\documentclass[structabstract]{aa}  

\usepackage{graphicx}
\usepackage{natbib}
\usepackage{morefloats}

\bibpunct{(}{)}{;}{a}{}{,} 
\usepackage{txfonts}
\begin{document}
 \title{Diffuse steep-spectrum sources from the 74~MHz VLSS survey} 
 

   \author{R.~J. van Weeren\inst{1}
         \and H.~J.~A. R\"ottgering\inst{1}
          \and M.~Br\"uggen \inst{2}
          }

   \institute{Leiden Observatory, Leiden University,
              P.O. Box 9513, NL-2300 RA Leiden, The Netherlands\\
              \email{rvweeren@strw.leidenuniv.nl}
                 \and Jacobs University Bremen, P.O. Box 750561, 28725 Bremen, Germany
                 }


 
\abstract
   {Galaxy clusters grow by a sequence of mergers with other clusters and galaxy groups. During these mergers, shocks and/or turbulence are created within the intracluster medium (ICM). In this process, particles could be accelerated to highly relativistic energies. The synchrotron radiation from these particles is observed in the form of radio relics and halos that are generally characterized by a steep radio spectral index. Shocks can also revive fossil radio plasma from a previous episode of AGN activity, creating a  so-called radio ``phoenix''. Here we present multi-frequency radio observations of  diffuse steep-spectrum radio sources selected from the 74~MHz VLSS survey. Previous Giant Metrewave Radio Telescope (GMRT) observations showed that some of these sources had filamentary and elongated morphologies, which are expected for radio relics.}
   {We attempt to understand the nature of diffuse steep-spectrum radio sources and characterize their spectral index and polarization properties.
   }
   {We carried out radio continuum observations at 325~MHz with the GMRT. Observations with the Very Large Array (VLA) and Westerbork Synthesis Radio Telescope (WSRT) were taken at 1.4~GHz in full polarization mode. Optical images around the radio sources were taken with the William Herschel and Isaac Newton Telescope (WHT, INT). 
   } 
   {Most of the sources in our sample consist of old radio plasma from AGNs located in small galaxy clusters. The sources can be classified as AGN relics or radio phoenices. The spectral indices across most of the radio sources display large variations.   
     }
{ We conclude that diffuse steep-spectrum radio sources are not only found in massive X-ray luminous galaxy clusters but also in smaller systems. Future low-frequency surveys will uncover large numbers of steep-spectrum radio relics related to previous episodes of AGN activity.
}
   \keywords{Radio Continuum: galaxies  -- Galaxies: active -- Clusters: general -- Cosmology: large-scale structure of Universe}

   \maketitle

\section{Introduction}
Studies of large-scale structure formation show that galaxy clusters grow 
through mergers with other clusters and galaxy groups, as well as through 
the continuous accretion of gas from the intergalactic medium (IGM). The 
baryonic content of clusters is mostly in the form of hot thermal gas 
visible at X-ray wavelengths. 
Several clusters also have a non-thermal component within the ICM, which is 
observable at radio wavelengths 
\citep[e.g.,][]{2008SSRv..134...93F, 2009A&A...507.1257G, 2000NewA....5..335G, 2010Sci...330..347V}. 
The idea is that shocks and/or turbulence generated during cluster merger 
events can accelerate particles to relativistic energies, and in the 
presence of a magnetic field, synchrotron radiation is emitted 
\citep[e.g.,][]{1998A&A...332..395E, 2001ApJ...562..233M, 2007MNRAS.375...77H, 2008MNRAS.391.1511H, 2008MNRAS.385.1242P, 2009MNRAS.393.1073B, 2010arXiv1006.3559S}. These radio sources, which trace the non-thermal component of the ICM, can be divided into several classes.  

Radio \emph{halos} are found at the center of merging galaxy clusters and have typical sizes of about a Mpc. They follow the X-ray emission from the thermal ICM and are mostly unpolarized, although there are some exceptions \citep[see][]{2005A&A...430L...5G, 2009A&A...503..707B}. Radio halos have been explained by turbulence injected by recent merger events. This injected turbulence might be capable of re-accelerating relativistic particles \citep[e.g., ][]{2001MNRAS.320..365B, 2001ApJ...557..560P}. Alternatively, the energetic electrons are secondary products of proton-proton collisions \citep[e.g.,][]{1980ApJ...239L..93D, 1999APh....12..169B, 2000A&A...362..151D}. The turbulent re-acceleration model currently seems to provide a better explanation of the occurrence of radio halos  \citep[e.g.,][]{2008Natur.455..944B}.

\emph{Mini-halos} (also called core-halo systems) are diffuse radio sources with sizes $\lesssim 500$~kpc located in relaxed galaxy clusters, in which diffuse emission surrounds the central cluster galaxy \citep{2009A&A...499..679M, 2009A&A...499..371G, 2007A&A...470L..25G, 2004A&A...417....1G, 2003A&A...400..465B, 2002A&A...386..456G, 1992ApJ...388L..49B}. 

Radio \emph{relics} are irregularly shaped radio sources with sizes ranging from 50~kpc to 2~Mpc, which can be divided into three groups \citep{2004rcfg.proc..335K}. \emph{Radio gischt} are large elongated, often Mpc-sized, radio sources located at the periphery of merging clusters. They probably trace shock fronts in which particles are accelerated via the diffusive shock acceleration
mechanism \citep[DSA;][]{1977DoSSR.234R1306K, 1977ICRC...11..132A, 1978MNRAS.182..147B, 1978MNRAS.182..443B, 1978ApJ...221L..29B, 1983RPPh...46..973D, 1987PhR...154....1B, 1991SSRv...58..259J, 2001RPPh...64..429M}. Among these are rare double-relics that have two relics located  on both sides of the  cluster center \citep[e.g., ][]{2009A&A...494..429B, 2009A&A...506.1083V, 2007A&A...463..937V, 2006Sci...314..791B, 1997MNRAS.290..577R, 2010Sci...330..347V, 2010arXiv1011.4985B}. According to DSA theory, the integrated radio spectrum should be a single power-law. \emph{Radio phoenices}  and \emph{AGN relics} are both related to radio galaxies. AGN relics are associated with extinct or dying radio galaxies. The radio plasma has a steep curved spectrum\footnote{$F_{\nu} \propto \nu^{\alpha}$, where $\alpha$ is the spectral index} due to synchrotron and inverse Compton (IC) losses. Fossil radio plasma from a previous episode of AGN activity can also be compressed by a merger shock wave producing a radio phoenix \citep{2001A&A...366...26E, 2002MNRAS.331.1011E}, these sources again having steeply curved radio spectra. Proposed examples of these are those found by \cite{2001AJ....122.1172S}. 

In \cite{2009A&A...508...75V}, we presented observations 
of a sample of 26 diffuse steep-spectrum sources, 
with $\alpha \le -1.15$, selected from the 1.4~GHz NVSS \citep{1998AJ....115.1693C} and 74~MHz VLSS \citep{2007AJ....134.1245C} surveys. These sources were either resolved out in the VLA B-array 1.4~GHz snapshot observations or  
1.4~GHz FIRST survey \citep{1995ApJ...450..559B}. GMRT 610~MHz 
observations of these 26 sources detected one distant powerful radio 
halo with a radio relic \citep{2009A&A...505..991V}, five radio relics 
including two radio phoenices, and one possible mini-halo. The remaining 
sources were classified as radio sources directly related to AGN activity. 
The spectral indices of the radio relics in the sample are generally steeper than 
most previously known relics. By complementing our observations with results  
for other relics from the literature, we found that the larger relics generally 
have flatter spectra and are located farther away from the cluster center. 
This is in line with predictions from shock statistics derived from 
cosmological simulations \citep{2008MNRAS.391.1511H, 2008ApJ...689.1063S, 2009MNRAS.393.1073B}. 
In these simulations, it is found that larger shock waves 
occur mainly in lower-density regions and have larger Mach 
numbers, and consequently shallower spectra. On the other hand, 
smaller shock waves are more likely to be found in cluster 
centers and have lower Mach numbers, thus steeper spectra. 

In this paper, we present follow-up radio observations of six 
sources (i.e., those most likely related to radio relics 
and halos) found in \cite{2009A&A...508...75V}. GMRT 325~MHz 
as well as VLA and WSRT 1.4~GHz observations were obtained to create 
spectral index and polarization maps. Optical images were acquired with 
the 4.2m~WHT and 2.5m~INT telescopes at the 
position of the radio sources to search for optical counterparts 
and identify galaxy clusters associated with the radio sources.

The layout of this paper is as follows. In Sect.~\ref{sec:obs-reduction},  
we present an overview of the observations and data reduction. 
In Sect.~\ref{sec:results}, we present the radio and spectral maps as 
well as optical images around the radio sources. In Sect.~\ref{sec:other},  
we show additional optical images around five slightly more compact radio 
sources (also from the sample of 26 sources) to search for optical counterparts. 
These sources are not located in nearby galaxy clusters and their nature 
remains unclear. We end with a discussion and conclusions in Sects.~\ref{sec:discussion} and~\ref{sec:conclusion}.

Throughout this paper, we assume a $\Lambda$CDM cosmology with $H_{0} = 71$ km s$^{-1}$ Mpc$^{-1}$, $\Omega_{m} = 0.3$, and $\Omega_{\Lambda} = 0.7$. All images are in the J2000 coordinate system.

\section{Observations \& data reduction}
\label{sec:obs-reduction}

\subsection{GMRT 325~MHz observations}
Radio continuum observations with the GMRT at 325~MHz were carried out on 14, 15, and 17~May, 2009. Both upper (USB) and lower (LSB) sidebands (IFs, which included RR and LL polarizations) were recorded with a total bandwidth of 32~MHz. The observations were carried out in spectral line mode with 128~channels per IF to facilitate the removal of radio frequency interference (RFI) and reduce the effect of bandwidth smearing. The integration time per visibility was 8~sec. Each source was observed for about 4 hrs in total. The data were reduced with the NRAO Astronomical Image Processing System (AIPS) package. 

The data was visually inspected for the presence of RFI, which was subsequently removed (i.e., ``flagged''). We carried out an amplitude and phase calibration on the flux and bandpass calibrators 3C147 and 3C286 on a timescale of 8~sec. For this, we chose three neighboring frequency channels free of RFI. These gain solutions were applied before determining the bandpass response of the antennas. This assures that any amplitude and/or phase variations during the scans on the calibrators are corrected before determining the bandpass solutions. At higher frequencies (e.g., 1.4~GHz),  both amplitude and phases are assumed to be constant during bandpass calibration. However, for the GMRT observing at low frequencies, this  assumption is not always valid and can affect the quality of the bandpass solutions as well as the determination of the flux scale.

After correcting for the bandpass response, both the amplitude 
and phase solutions for both primary and secondary 
calibrators were determined but in this case using the full 
channel range. 
The fluxes of the primary calibrators were set according to 
the \cite{perleyandtaylor} extension to the \cite{1977A&A....61...99B} 
scale. 
The flux densities for the secondary calibrators were 
bootstrapped from the primary calibrators. The  
amplitude and phase solutions were interpolated 
and applied to the target sources. Some targets were 
observed over multiple days (observing runs), the 
resulting different data sets were combined with the 
AIPS task `DBCON'.

For each of the target sources, we created a model of the surrounding 
field using the NVSS survey with a spectral index scaling of $-0.7$. We carried out a phase-only 
self-calibration against this model to improve the astrometric accuracy.  
This was followed by several rounds of phase self-calibration and two 
final rounds of amplitude and phase self-calibration. To produce the images,  
we used the polyhedron method \citep{1989ASPC....6..259P, 1992A&A...261..353C} 
to minimize the effects of non-coplanar baselines. 
The model was then subtracted from the data,  a step that facilitated the removal 
of additional RFI or baselines with problems. Final images were made 
using robust weighting \citep[robust = 0.5,][]{briggs_phd}.
 Images were cleaned using the automatic clean-box windowing algorithm 
in AIPS and cleaned down to $2$ times the rms noise level ($2\sigma_{\mathrm{rms}}$) 
within the clean boxes. The final images were corrected for the primary beam 
response\footnote{http://gmrt.ncra.tifr.res.in/gmrt\_hpage/Users/doc/manual/
 
 UsersManual/node27.html}. The uncertainty in the calibration of the 
absolute flux-scale is in the range $5-10\%$, see \cite{2004ApJ...612..974C}. 
The resulting noise levels and beam sizes are shown in Table~\ref{tab:gmrtobservations}.

\begin{table}
\begin{center}
\caption{GMRT 325~MHz observations}
\begin{tabular}{lll}
\hline
\hline
& rms noise  & beam size  \\
& $\mu$Jy~beam$^{-1}$ & arcsec \\
\hline
VLSS J1431.8+1331 & 178 &$11.8\arcsec \times 7.8\arcsec$ \\
VLSS J1133.7+2324 &  132&$10.4\arcsec \times 7.7\arcsec$  \\
Abell~2048                   & 248 &$9.7\arcsec \times 9.6\arcsec$  \\
24P73                             &162  &$13.6\arcsec \times 9.0\arcsec$\\
VLSS J0004.9$-$3457& 309& $13.5\arcsec \times 11.0\arcsec$ \\
VLSS J0915.7+2511   & 412 &$14.7\arcsec \times 8.1\arcsec$\\
\hline
\end{tabular}
\label{tab:gmrtobservations}
\end{center}
\end{table}

Radio observations at $610$~MHz were taken with the GMRT in February and November 2008 of the sources in Table~\ref{tab:gmrtobservations}. The reduction of these observations is similar to the GMRT 325~MHz data and is described in more detail in \cite{2009A&A...508...75V}. We used these images to create the spectral index maps.

\begin{table*}
\begin{center}
\caption{VLA 1.4~GHz observations}
\begin{tabular}{lllll}
\hline
\hline
& VLSS J1133.7+2324 & MaxBCG J217.95869+13.53470 & VLSS J0004.9$-$3457 & Abell 2048\\
\hline
Frequency bands (IFs)             & 1385, 1465~MHz & 1385, 1465~MHz& 1385, 1465~MHz & 1385, 1465~MHz\\
Bandwidth		  & $2\times50$~MHz& $2\times50$~MHz & $2\times50$~MHz& $2\times50$~MHz \\
Polarization			& RR, LL, RL, LR&RR, LL, RL, LR & RR, LL, RL, LR&RR, LL, RL, LR\\
Observation dates				& 4 Nov 2008, 17 Apr 2009, & 2 Nov 2008, 18 Apr 2009, & 11 Jun 2009&15 Jul 2009 \\ 
                                                                  &10 Aug 2009   &  31 Jul 2009 &&\\
Project code &AV305, AV312&AV305, AV312&AV312&AV312 \\
Integration time             & 3.3~s &3.3~s& 3.3~s &3.3~s \\
Total on-source time		&5.0~hr, 6.8~hr, 3.9~hr & 4.9~hr, 6.8~hr, 3.9~hr& 4.0~hr& 4.0~hr\\
VLA configuration   &A+B+C&A+B+C&CnB&C\\
Beam size			& $1.3\arcsec \times 1.4\arcsec^{a}$, $6.7\arcsec \times 6.7\arcsec^{b}$ & $1.6\arcsec \times 1.5\arcsec^{a}$, $6.4\arcsec \times 5.3\arcsec^{b}$ &$19.8\arcsec \times 10.3\arcsec$ & $13.0\arcsec \times 12.4\arcsec$\\
Rms noise ($\sigma_{\mathrm{rms}}$)	& 14$^{a}$,  19$^{b}$ $\mu$Jy beam$^{-1}$ & 15$^{a}$, 17$^{b}$ $\mu$Jy beam$^{-1}$&49 $\mu$Jy beam$^{-1}$&88 $\mu$Jy beam$^{-1}$ \\
\hline
\hline
\end{tabular}
\label{tab:vlaobservations}
\end{center}
$^{a}$ Briggs weighting (robust = $-1.0$)\\
$^{b}$ natural weighting
\end{table*}
\subsection{VLA 1.4~GHz observations}
\label{sec:vlaobs}
We carried out L-band observations of four sources with the VLA (see Table~\ref{tab:vlaobservations}). The observations were taken in standard continuum mode with two IFs, each having a bandwidth of 50~MHz recording all polarization products (RR, LL, RL, and LR). Gain solutions were determined for the calibrator sources and transferred to the target sources. The fluxes for the primary calibrators were set according to the \cite{perleyandtaylor} extension to the \cite{1977A&A....61...99B} scale.  The effective feed polarization parameters (the leakage terms or D-terms) were found by observing the phase calibrator over a wide range of parallactic angles  and simultaneously solving for the unknown polarization properties of the source.  The polarization angles were set using the polarized sources 3C286 and 3C138. For the R-L phase difference, we assumed values of  $-66.0$ and $15.0$~deg for 3C286 and 3C138, respectively. Stokes Q and U images were compiled for each source. From the Stokes Q and U images, the polarization angles ($\Psi$) were determined ($\Psi = \frac{1}{2} \arctan{(U/Q})$). Total polarized intensity ($P$) images were also made ($P = \sqrt{Q^2 + U^2}$). The polarization fraction were found by dividing the total polarized intensity by the total intensity (Stokes I) image ($ \sqrt{Q^2 + U^2}$/I).

\subsection{WSRT $1.3-1.7$~GHz observations of 24P73}

\begin{table*}
\begin{center}
\caption{WSRT observations}
\begin{tabular}{ll}
\hline
\hline
Frequency bands 21 cm (IFs)             & 1311, 1330, 1350, 1370, 1392, 1410, 1432, 1450~MHz  \\
Frequency bands 18 cm (IFs)            &  1650, 1668, 1686, 1704, 1722, 1740, 1758, 1776~MHz\\
Bandwidth per IF                                 & 20~MHz \\
Number of channels per IF                     & 64 \\
Channel width                                      & 312.5~kHz\\
Polarization			                 & XX, YY, XY, XY \\
Observation dates		                &  11, 17, and 18 March, 2009 \\
Integration time                                & 30~s \\
Total on-source time		             & $6.5$~hr 21cm + $6.5$~hr 18cm\\
Beam size			&  $19.0\arcsec \times 16.5\arcsec$  \\
Rms noise ($\sigma_{\mathrm{rms}}$) & 37~$\mu$Jy~beam$^{-1}$ \\
\hline
\hline
\end{tabular}
\label{tab:wsrtobservations}
\end{center}
\end{table*}

WSRT observations were taken of a single source   
(\object{24P73}) not included in the VLA observations. Every 5~min, the frequency setup was changed within the L-band, alternating between the 21cm and 18cm setups. Both of these frequency setups have 160~MHz bandwidth divided over 8~IFs, each having 20~MHz bandwidth. The data were recorded in spectral line mode with 64 spectral channels per IF in 4 polarizations. The observations were carried out in three runs on 11, 17, and 18 March, 2009 resulting in a more or less complete 12-hour synthesis run, see Table~\ref{tab:wsrtobservations}. 

The data were partly calibrated using the CASA (formerly AIPS++)\footnote{http://casa.nrao.edu/} package. The L-band receivers of the WSRT telescopes have linearly polarized feeds\footnote{The WSRT records $\rm{XX}=\rm{I}-\rm{Q}$, $\rm{YY}=\rm{I}-\rm{Q}$, $\rm{XY} = -\rm{U }+ i\rm{V}$, and $\rm{XY} = -\rm{U}-i\rm{V}$, where I, Q, U, and V are the Stokes parameters.}. The leakage terms (D-terms) for the WSRT are frequency dependent \citep[e.g.,][]{2008A&A...489...69B}. As a first step, we flagged the autocorrelations, and removed any obvious RFI  and corrupted data. Time ranges of antennas affected by shadowing were also taken out. Bandpass and gain solutions were determined using observations of two standard calibrators, both at the start and end of an observing run. The fluxes for the calibrators were set according to the \cite{perleyandtaylor} extension to the \cite{1977A&A....61...99B} scale. We used both polarized (3C138 or 3C286) and unpolarized (CTD93, 3C48) calibrator sources. The bandpass and gain solutions were applied and the data were calibrated for the leakage terms using the unpolarized calibrator source. The polarization angles were set using the polarized calibrators sources, the angles being the same as in Sect.~\ref{sec:vlaobs}. The data was then exported into AIPS for two rounds of phase only and two rounds of amplitude and phase self-calibration (separately for each IF). The solutions for the amplitude and phase self-calibrations were determined by combing both XX and YY polarizations as Stokes Q is not necessarily zero.

The images for each IF were cleaned to about $2\sigma_{\mathrm{rms}}$ using clean boxes. The images for each IF were combined into a deep image by convolving the images of the individuals IFs to a common resolution and using a spectral index scaling of $-1$.

\subsection{Optical WHT \& INT imaging}
Optical images around the radio sources were made using the PFIP camera on the 4.2m WHT telescope and the WFC camera on the INT. The observations were carried out between 15 and 19 April, 2009 (WHT) and $1 - 8$ October, 2009 (INT). The field of view was $16\arcmin \times 16\arcmin$ for the PFIP and $34\arcmin \times 34\arcmin$ for the WFC camera. 
The seeing varied between 0.6\arcsec~and 2.0\arcsec, but was mostly between 1.0\arcsec~and 1.5\arcsec. Most nights were photometric. The total integration time per target was about 1500~s for both V, R, and I bands for the WHT observations and 4000~s for the INT observations. The data were reduced with IRAF \citep{1986SPIE..627..733T, 1993ASPC...52..173T} and the \emph{mscred} package \citep{1998ASPC..145...53V}. All images were flat-fielded and bias-corrected. The I and R band images were fringe corrected. The individual exposures were averaged, with pixels being rejected above $3.0\sigma_{\mathrm{rms}}$ to remove cosmic rays and other artifacts.

Zero-points were determined using various observations of standard stars taken during the nights.  Images taken on non-photometric nights were scaled such that the flux of a few targets within the field of view agreed with that of the images taken on photometric nights.

\section{Results}
A list of the observed sources with their integrated fluxes 
and redshifts is given in Table~\ref{tab:results}. For 
sources without a spectroscopic redshift, we used the Hubble-K 
or Hubble-R relations to estimate the redshift \citep{2003MNRAS.339..173W, 1996MNRAS.279.1294S, 2007A&A...464..879D}. 
Spectral index maps were also made for the radio sources. Only common 
UV-ranges were used to minimize errors due to differences in the UV-coverage 
and the individual maps were convolved to the same resolution. For sources 
with high-resolution ($\lesssim10$\arcsec) 1.4~GHz observations, we created 
a high-resolution spectral index map between 610 and 1425~MHz. 
 A low-frequency spectral index map between 325 and 610~MHz 
was also made for most sources. Pixels below $4\sigma_{\mathrm{rms}}$ at 
either one of the two frequency maps were blanked. Low-resolution spectral 
index maps between 325, 610, and 1425~MHz were created to map the spectral 
index in low signal-to-noise ratio (SNR) regions. We fitted a single (power-law) spectral 
index through the three frequencies, pixels below $2.5\sigma_{\mathrm{rms}}$ being neglected.

For two sources, the SNR was high enough to create spectral curvature maps. 
The spectral curvature we defined as $\alpha_{325-610} - \alpha_{610-1425}$. 
Pixels with a spectral index error larger than $0.06$ were blanked in the 
spectral curvature maps. The errors in the spectral index map are based on 
the noise levels ($\mathrm{rms}$) of the individual images. 
In the following subsections, the radio and spectral index maps are presented 
together with optical overlays.
\label{sec:results}

\begin{table*}
\begin{center}
\caption{Source list \& properties}
\begin{tabular}{lllllllll}
\hline
\hline
source/cluster& $z$ &S$_{325}$ & S$_{1425}$   & $\alpha_{74-1400}$& curvature&LLS & classification$^{h}$ & \\
 & & Jy & mJy & &$\alpha_{74-610} - \alpha_{610-1425}$ & kpc\\
\hline
\object{VLSS J1431.8+1331} &  0.1599 & $0.373 \pm 0.040$ & $14.6 \pm 1.1$&$-2.03 \pm 0.05$ & 1.03$^{f}$&125& AGN + AGNR or PHNX \\
\object{VLSS J1133.7+2324} &  $0.61 \pm 0.16^{g}$& $0.273 \pm 0.028$ & $12.3 \pm 1.1$ & $-1.69 \pm 0.06$ & 0.96 & 570$^{g}$ & AGNR or DSAR\\
Abell~2048                   & 0.0972  & $0.559 \pm 0.061$& $18.9 \pm 4.3$&$-1.50 \pm 0.05$& 1.6 & 310 & PHNX\\
24P73                             & $0.15 \pm 0.1^{b,g}$ & $0.307 \pm 0.033$ & $12.0 \pm 3.0$ &$-2.20\pm0.06$& 2.0& 270 & PHNX \\ 
\object{VLSS J0004.9$-$3457}& $0.3 \pm 0.1^{e}$&  $0.417 \pm 0.046$ & $32.2 \pm 1.9$ &$-1.40 \pm0.04$& 0.30$^{f}$ & 200 & AGN (or MH + DSAR?) \\
\object{VLSS J0915.7+2511}   & 0.324 & $0.417 \pm 0.046$ & $24.7 \pm 1.5^{a}$ & $-1.52 \pm 0.04$ & 1.02 & 190 & AGNR or PHNX\\
\object{VLSS J1117.1+7003}     & $0.8 \pm 0.4^{d}$&$0.030\pm 0.006^{c}$ & $2.9 \pm 0.5^{a}$& $-1.87 \pm 0.07$ & 0.0 & 130 & AGN?  \\
\object{VLSS J2043.9$-$1118} &$0.5 \pm 0.3^{d}$& \ldots&   $7.7 \pm0.6^{a}$ &$-1.74 \pm 0.05$&0.84& 250 & AGN (MH? + DSAR?)\\ 
\object{VLSS J0516.2+0103}    &$1.2\pm0.7^{d}$& \ldots &   $4.3 \pm0.4^{a}$  &$-1.73\pm0.06$& 0.0& 290 & AGN (or MH?)\\
\object{VLSS J2209.5+1546}    &$1.1 \pm 0.7^{d}$& \ldots &  $7.0 \pm0.9^{a}$ &$-1.56\pm0.07$& 0.59 & 500 & AGN? \\ 
\object{VLSS J2241.3$-$1626} &$0.5 \pm 0.3^{d}$& \ldots &  $14.6 \pm1.1^{a}$ &$-1.44\pm0.06$&0.0 & 290 & AGN?\\
\hline
\hline
\end{tabular}
\label{tab:results}
\end{center}
$^{a}$ flux from NVSS \citep{1998AJ....115.1693C}\\
$^{b}$ identification of the cD galaxy uncertain \\
$^{c}$ flux from WENSS \citep{1997A&AS..124..259R}\\
$^{d}$ redshift estimated using the fitted Hubble-R relation from \cite{2007A&A...464..879D}, since it is unclear whether there is a common underlying population of massive elliptical galaxies for extended steep-spectrum radio sources we have taken the 3C Hubble-R relation from \cite{1996MNRAS.279.1294S} as an upper limit for the redshift (i.e., the 3C galaxies are about 1 mag brighter at the same redshift) \\
$^{e}$ redshift estimated using the fitted Hubble-K relation from \cite{2003MNRAS.339..173W}\\
$^{f}$ varies across the source\\
$^{g}$ association with cluster uncertain\\
$^{h}$ PHNX = radio phoenix, AGNR = AGN relic, MH = radio mini-halo, DSAR = relic tracing shock wave with DSA\\
\end{table*}

\subsection{VLSS J1133.7+2324, \object{7C 1131+2341}} 
VLSS~J1133.7+2324 is a filamentary radio source, possibly associated with a galaxy cluster at $z=0.61$ \citep{2009A&A...508...75V}. To the west of the filamentary  source, our 610~MHz image detected diffuse emission associated with the foreground galaxy \object{UGC~6544} located at $z=0.02385$ \citep{1997AJ....113.1197H}.

\begin{figure*}
\begin{center}
\includegraphics[angle =90, trim =0cm 0cm 0cm 0cm,width=0.45\textwidth]{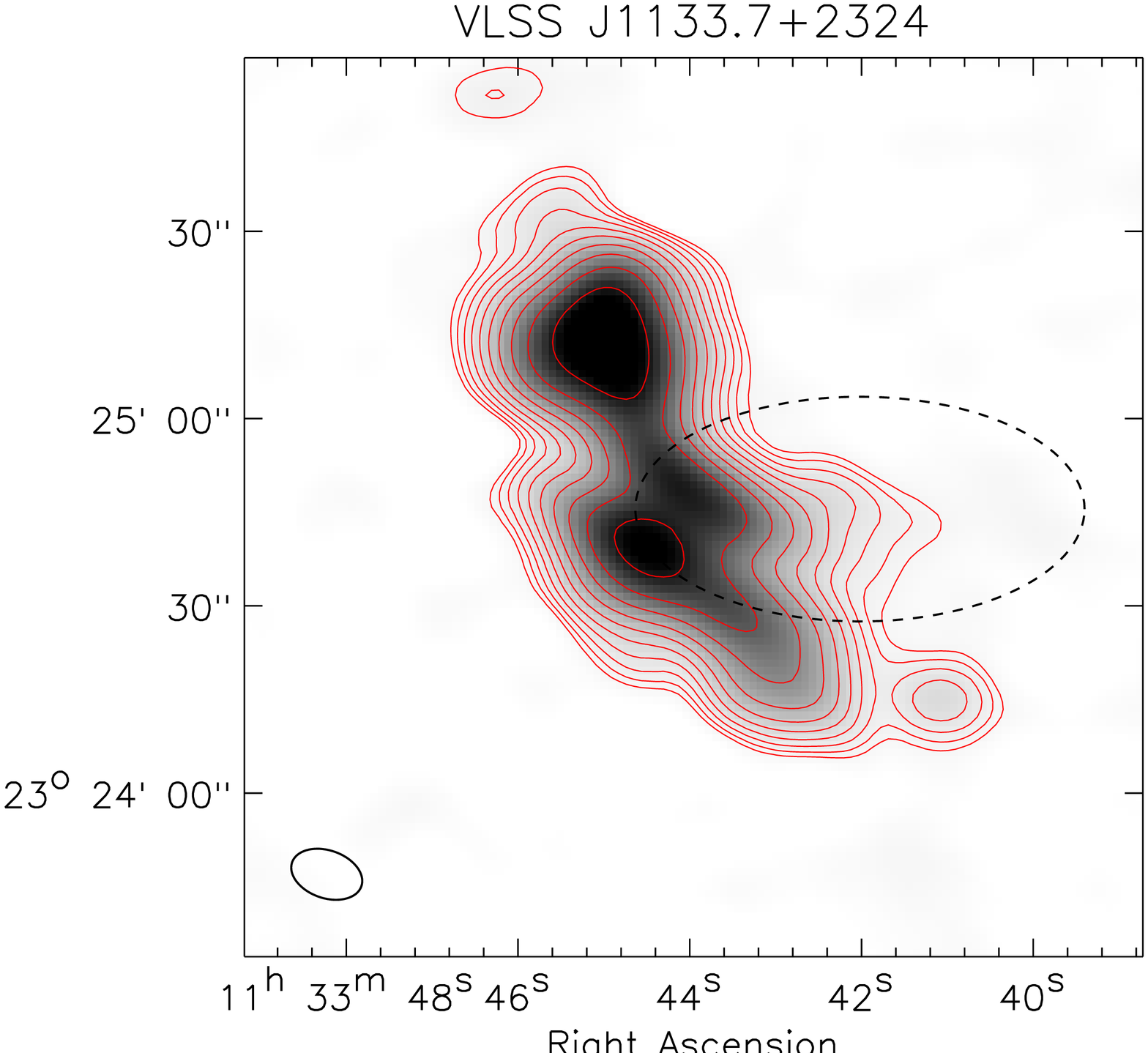}
\includegraphics[angle =90, trim =0cm 0cm 0cm 0cm,width=0.45\textwidth]{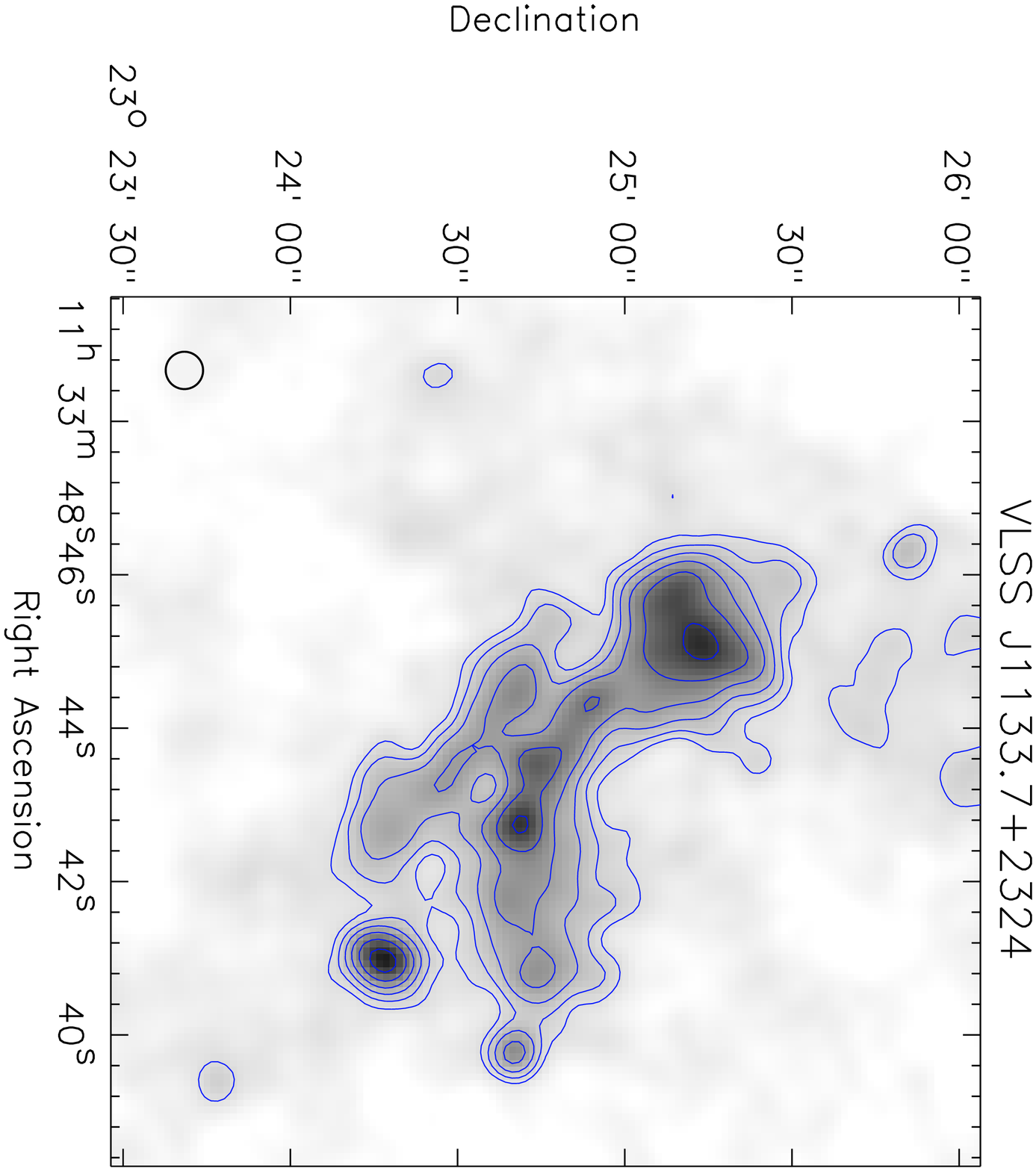}
\end{center}
\caption{Left: GMRT 325 MHz map. Contour levels 
are drawn at $\sqrt{[1, 2, 4, 8, \ldots]} \times 4\sigma_{\mathrm{rms}}$. 
The position of is UGC~6544 is indicated 
by the dashed ellipse. Right: VLA 1425 MHz map. Contour 
levels are drawn as in the left panel. }
\label{fig:gmrt325_5}
\end{figure*}

\begin{figure}
\begin{center}
\includegraphics[angle =90, trim =0cm 0cm 0cm 0cm,width=0.5\textwidth]{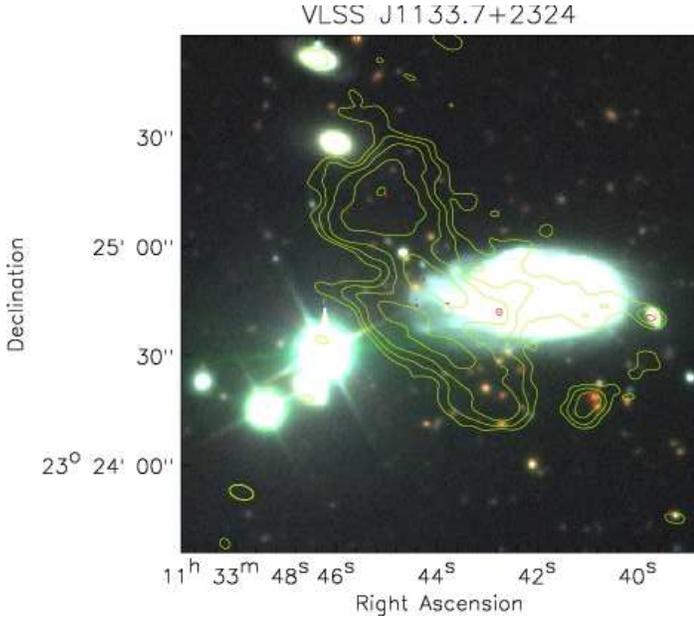}
\end{center}
\caption{Optical WHT color image for VLSS~J1133.7+2324. GMRT 610~MHz contours are overlaid in yellow. The beam size is $6.6\arcsec \times 3.9\arcsec$. The VLA 1425 MHz high-resolution map is overlaid with red contours. The beam size is $1.36\arcsec \times 1.32\arcsec$~and the image has a noise of 14~$\mu$Jy~beam$^{-1}$. Contour levels are drawn at $[1, 2, 4, 8, \ldots] \times 4\sigma_{\mathrm{rms}}$. }
\label{fig:s5_optical}
\end{figure}

The 325~MHz image (see Fig.~\ref{fig:gmrt325_5} left panel), 
is similar to our previous 610~MHz image. 
In the 325~MHz image, the southern part of the 
filamentary source is significantly brighter than  
the northern part, while  
 little emission from UGC~6544 is detected at 325~MHz. 
The VLA 1.4~GHz image made with natural weighting is 
shown in Fig.~\ref{fig:gmrt325_5} (right panel) and 
reveals much more emission from UGC~6544. The southern 
part of the filamentary source is quite faint at 1.4~GHz, 
confirming that it has a steep spectrum. A high-resolution VLA 
image, also at 1.4~GHz, is shown as an overlay in Fig.~\ref{fig:s5_optical}. 
This image reveals only three faint compact radio sources. This 
shows that the emission from the steep-spectrum source is truly diffuse, and 
cannot be attributed to the combined emission from compact sources.
We do not detect any polarized emission from the source at 1425~MHz. 
We place a $5\sigma$ upper limit of 5\% on the polarization fraction.

\begin{figure*}
\begin{center}
\includegraphics[angle =90, trim =0cm 0cm 0cm 0cm,width=0.45\textwidth]{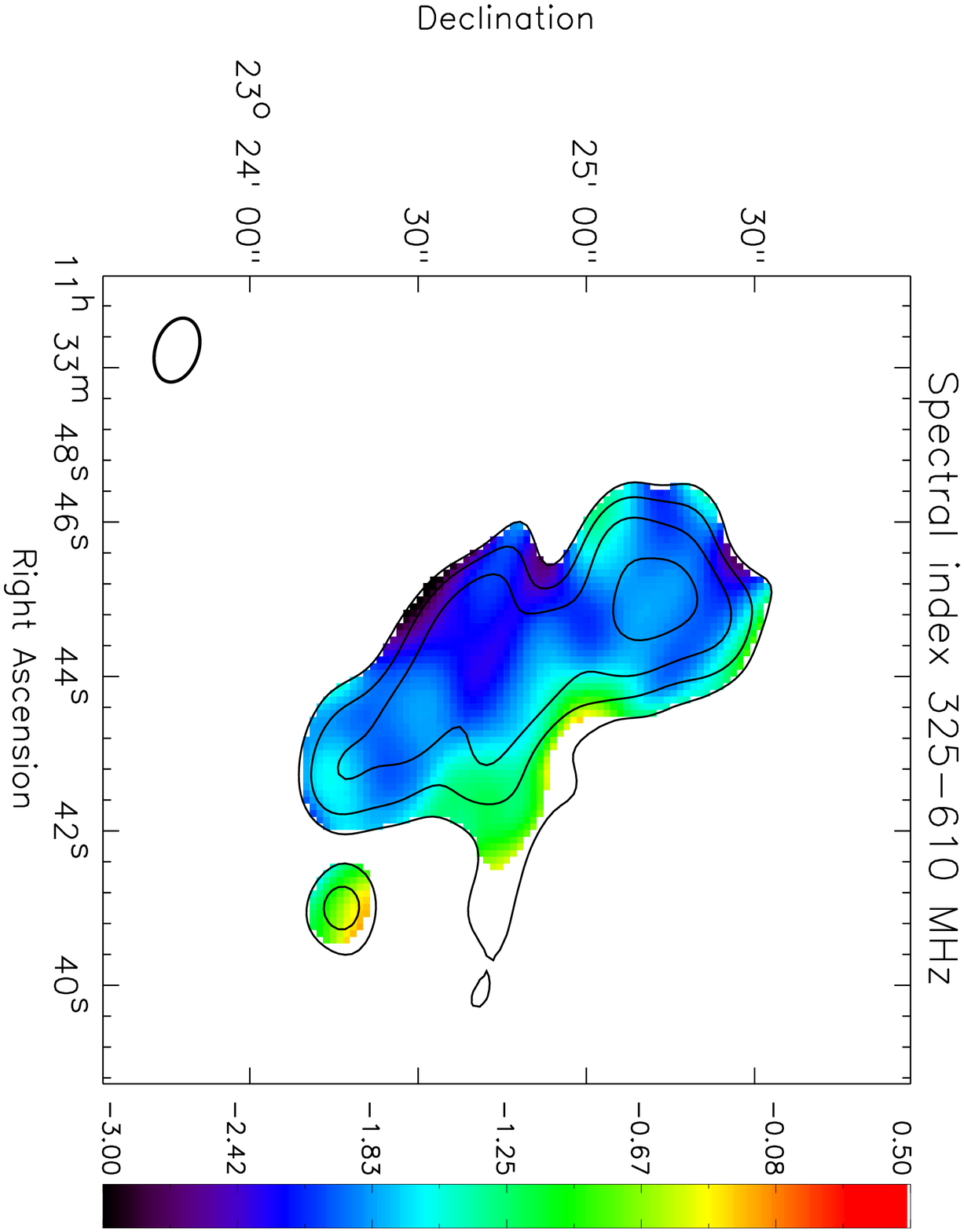}
\includegraphics[angle =90, trim =0cm 0cm 0cm 0cm,width=0.45\textwidth]{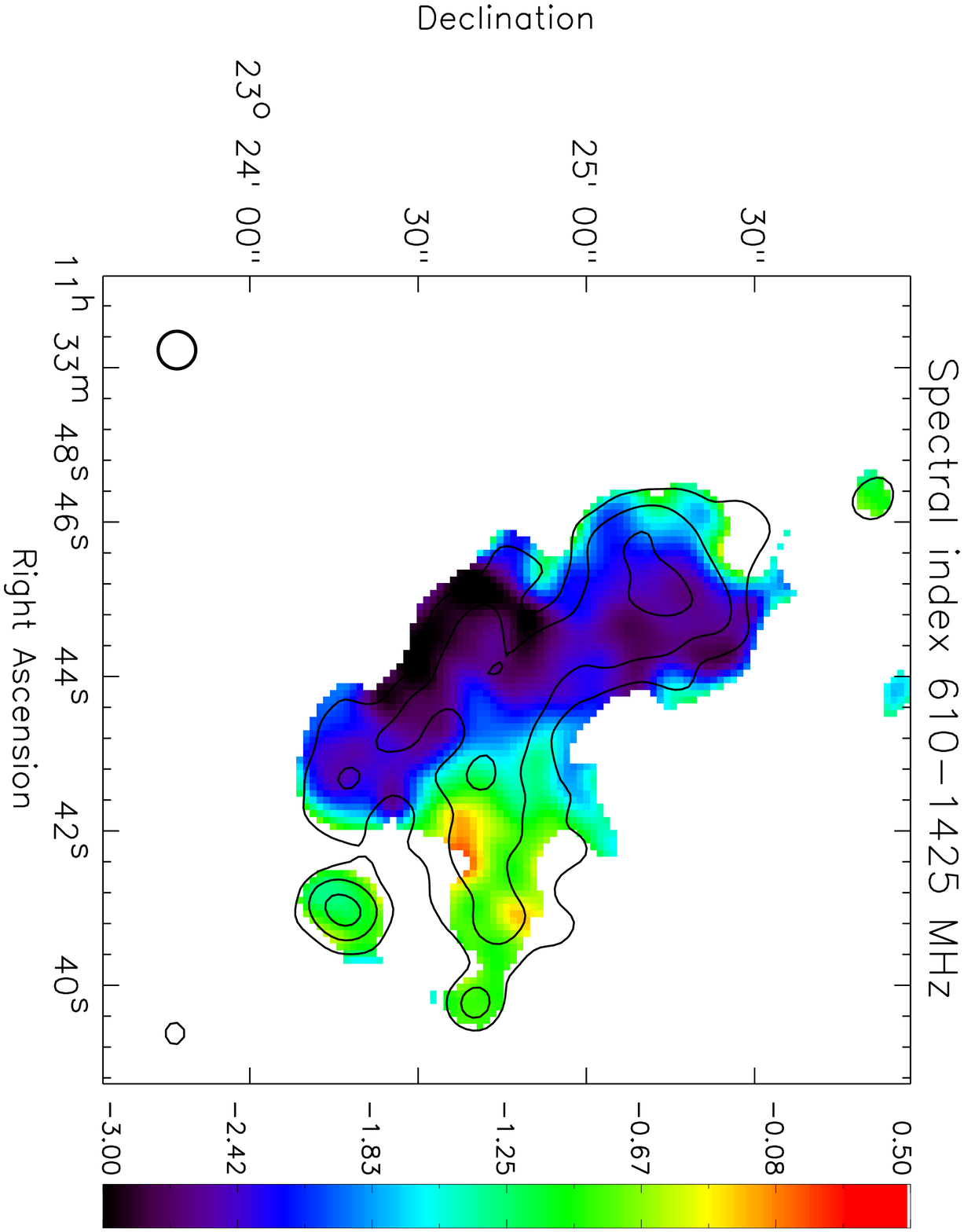}
\end{center}
\caption{Left: Spectral index map between 325 and 610 MHz at a resolution of $11.75\arcsec \times 7.65\arcsec$.  Contour levels are from the 325 MHz GMRT image  and drawn at levels of $[1, 2, 4, 8, \ldots]~\times 6\sigma_{\mathrm{rms}}$. Right: Spectral index between 610 and 1425~MHz. Contour levels are from the 1425~MHz VLA image and drawn at levels of $[1, 2, 4, 8,  \ldots]~\times 5\sigma_{\mathrm{rms}}$. The beam size is $6.7\arcsec \times 6.7\arcsec$. } 
\label{fig:spix5}
\end{figure*}

The spectral index map, between 325 and 610~MHz, 
is shown in Fig.~\ref{fig:spix5} (left panel). 
The spectral index of the southern part of the 
filamentary source is $-2.0$, while for the rest of the 
diffuse emission it is $\alpha \sim -1.7$. Towards UGC~6544,  
the spectral index flattens. The spectral index map between 
610 and 1425~MHz is shown in Fig.~\ref{fig:spix5} (right panel). 
Here the spectral index steepens to $\alpha \le -2.5$ for the 
southern part of the filamentary source. The spectral index of the foreground galaxy UGC~6544 is 
relatively flat with $\alpha\sim -0.5$.

VLSS~J1133.7+2324 was also observed by 
\cite{2009ApJ...698L.163D} at 1287~MHz with 
the GMRT and at 330~MHz with the VLA.  The 
reported integrated spectral indices were $-1.6 \pm 0.03$ 
between 74 and 328~MHz and $-1.9 \pm 0.08$ between 328 and 
1278~MHz. The integrated flux density was $151 \pm 12$~mJy 
at 328~MHz. We measure a flux of $273 \pm 28$~mJy at 325~MHz, 
which is significantly higher than the reported value from 
\citeauthor{2009ApJ...698L.163D}. This may be partly caused 
by the higher SNR of our image as we may pick up additional 
emission beyond what is visible in the \citeauthor{2009ApJ...698L.163D} 
image. Although, this cannot completely explain the difference in fluxes.

The source could be old radio plasma from a 
previous episode of AGN activity from \object{UGC~6544}, 
but UGC~6544 is a spiral galaxy, which normally do not host AGN. 
If the radio emission were explained by relic 
lobes, they would be expected to be located 
symmetrically with respect to the nucleus of 
the galaxy, which is not the case. The flat spectrum 
radio emission we detect from the galaxy is fully 
consistent with that predicted by the far-infrared 
radio correlation \citep{2009A&A...508...75V}.  

At the location of the filamentary radio source, 
we detected an overdensity of faint galaxies  
(see Fig.~\ref{fig:s5_optical}). These galaxies are 
partly hidden behind UGC~6544. The median SDSS 
photometric redshift is $0.61$ for these galaxies. 
To confirm the presence of a cluster, X-ray observations 
and/or spectroscopic redshifts of several galaxies are needed.

If this is indeed a distant galaxy cluster, the radio emission is 
very likely to be associated with the cluster. 
The source may then trace a shock wave in the cluster 
where particles are accelerated by the DSA mechanism. In 
that case the integrated radio spectrum should be a single 
power-law. Our flux measurements however indicate a slightly 
curved spectrum. A redshift of 0.61 would correspond to a 
physical size of about 500~kpc. The very steep and somewhat 
curved radio spectrum then suggest the source to be an AGN 
relic, rather than a radio phoenix because the source is 
quite large and the time to compress such a large radio ``ghost'' 
would remove most of the electrons responsible for the radio 
emission by radiative energy losses \citep{2006AJ....131.2900C}. 
Additional flux measurements at lower and/or higher frequencies  
will be needed to confirm whether the radio spectrum is indeed curved.

\subsection{VLSS~J1431.8+1331, MaxBCG~J217.95869+13.53470}
\begin{figure*}
\begin{center}
\includegraphics[angle =90, trim =0cm 0cm 0cm 0cm,width=0.32\textwidth]{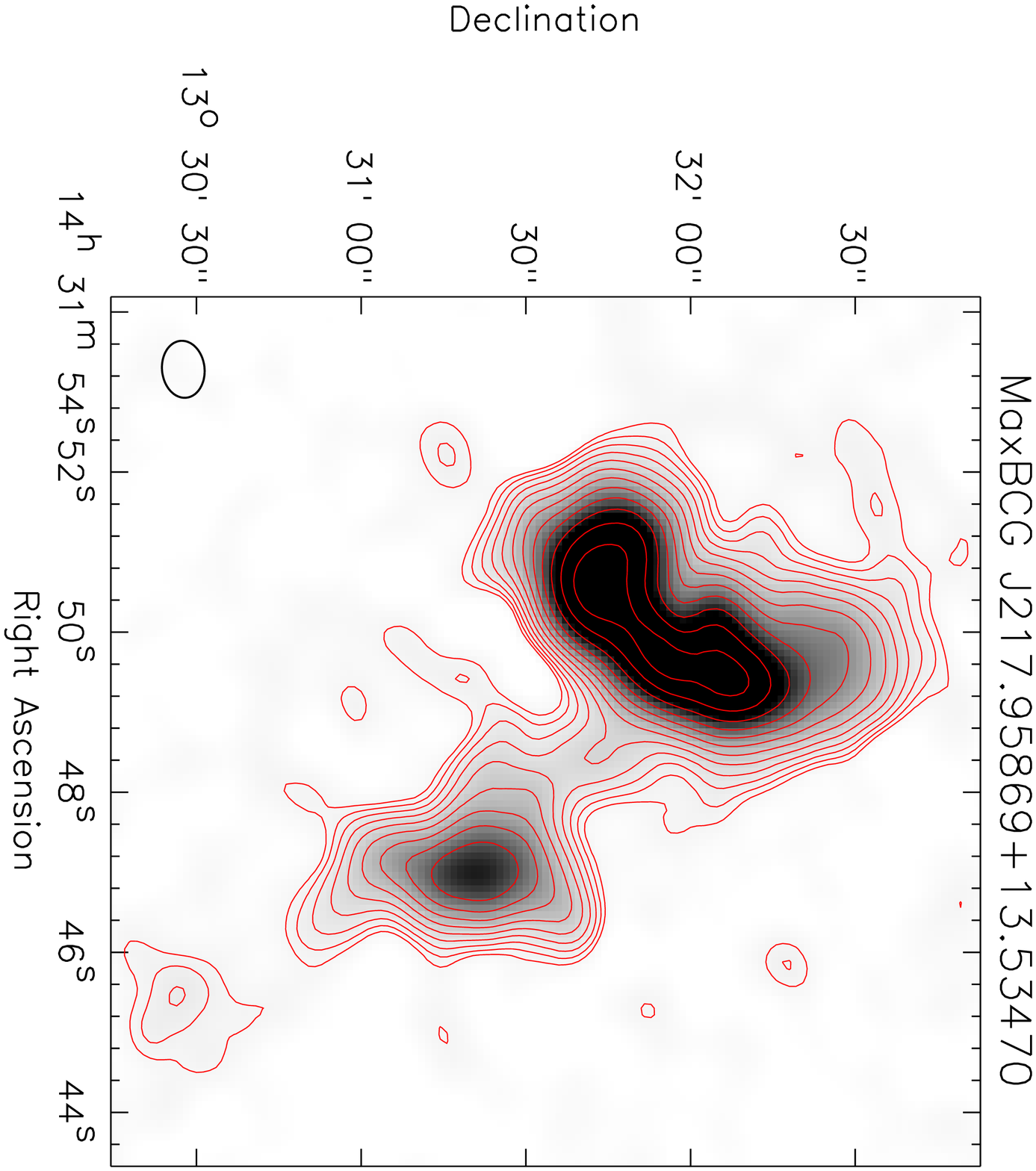}
\includegraphics[angle =90, trim =0cm 0cm 0cm 0cm,width=0.32\textwidth]{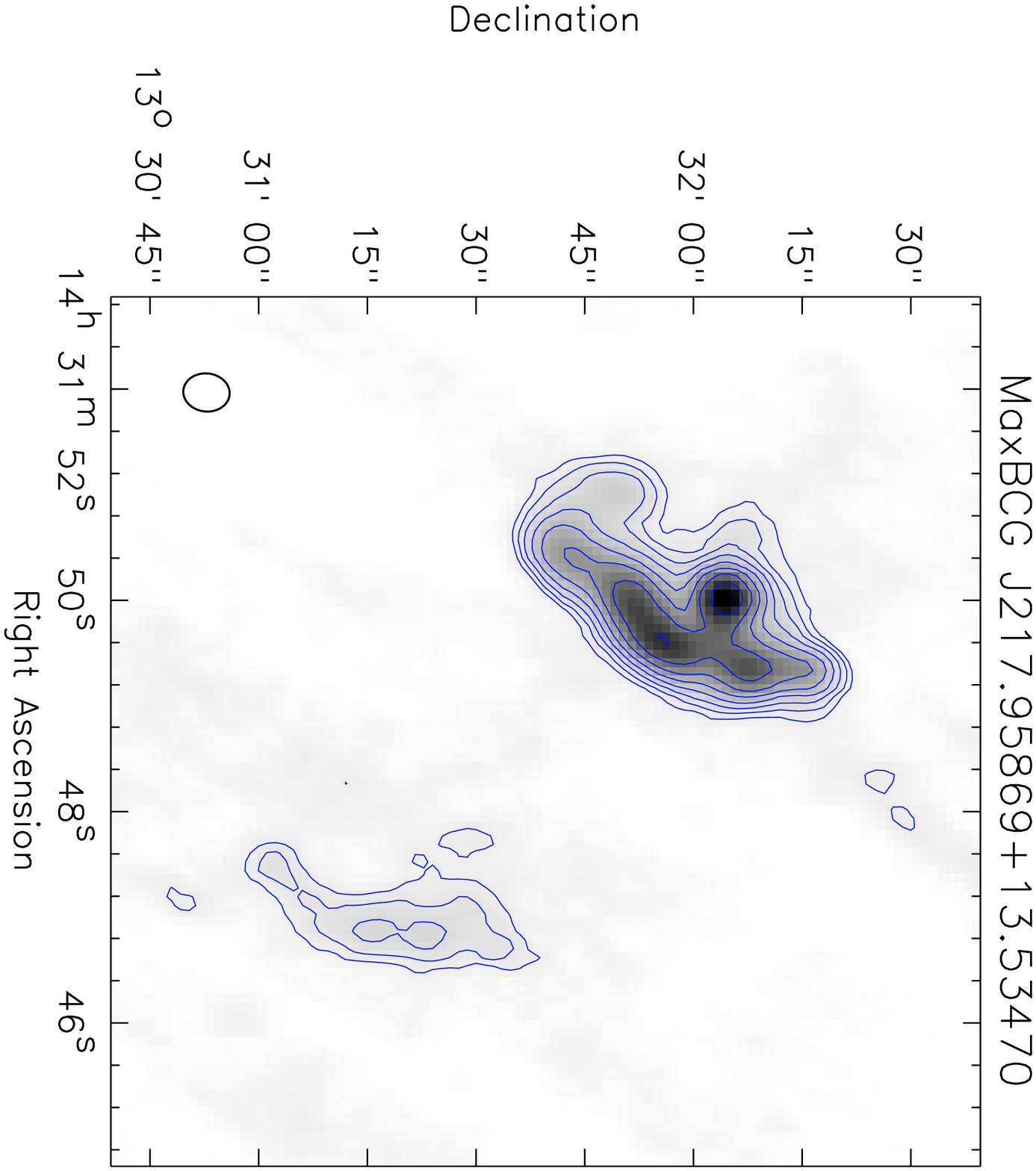}
\includegraphics[angle =90, trim =0cm 0cm 0cm 0cm,width=0.32\textwidth]{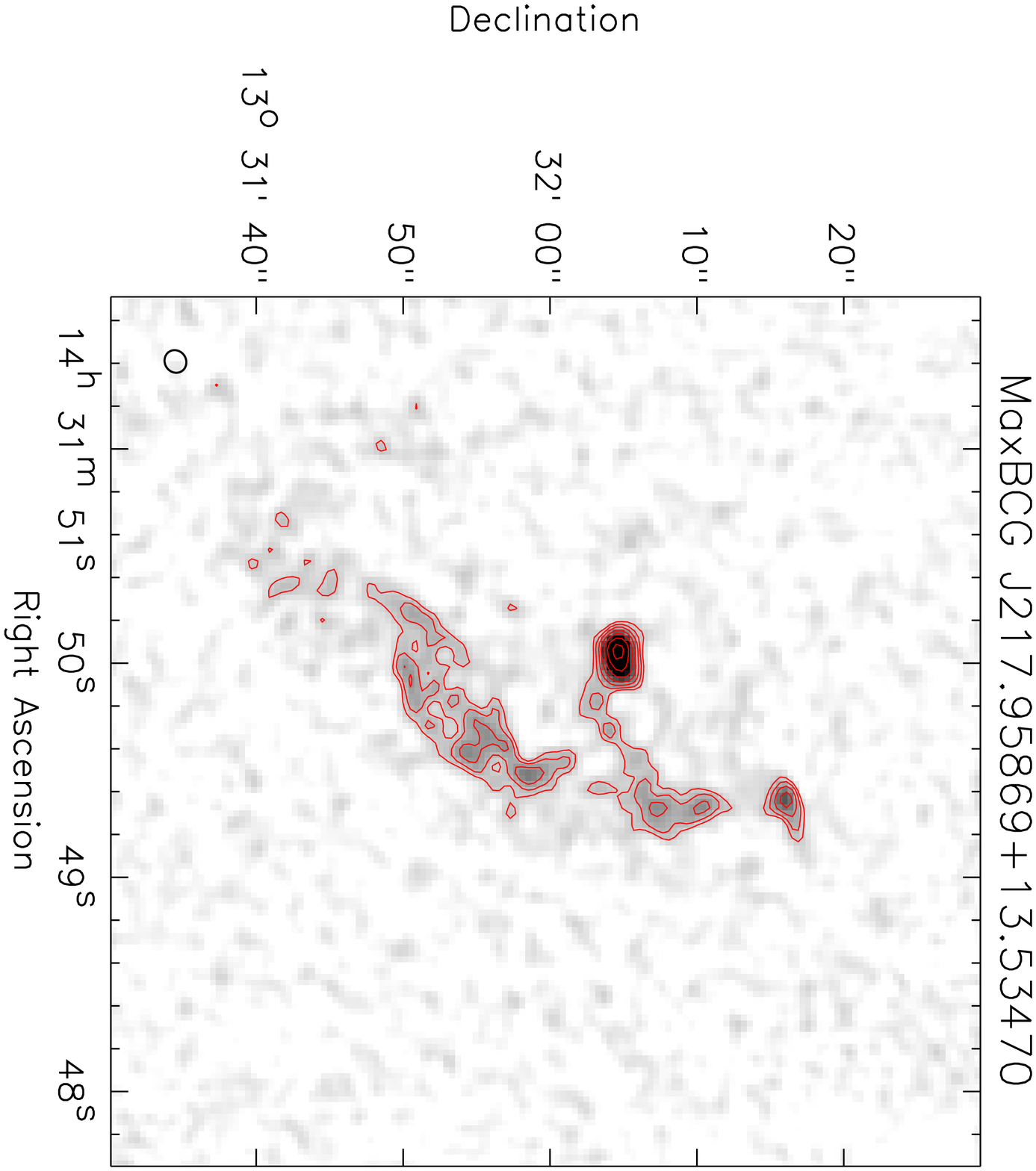}
\includegraphics[angle =90, trim =0cm 0cm 0cm 0cm,width=0.32\textwidth]{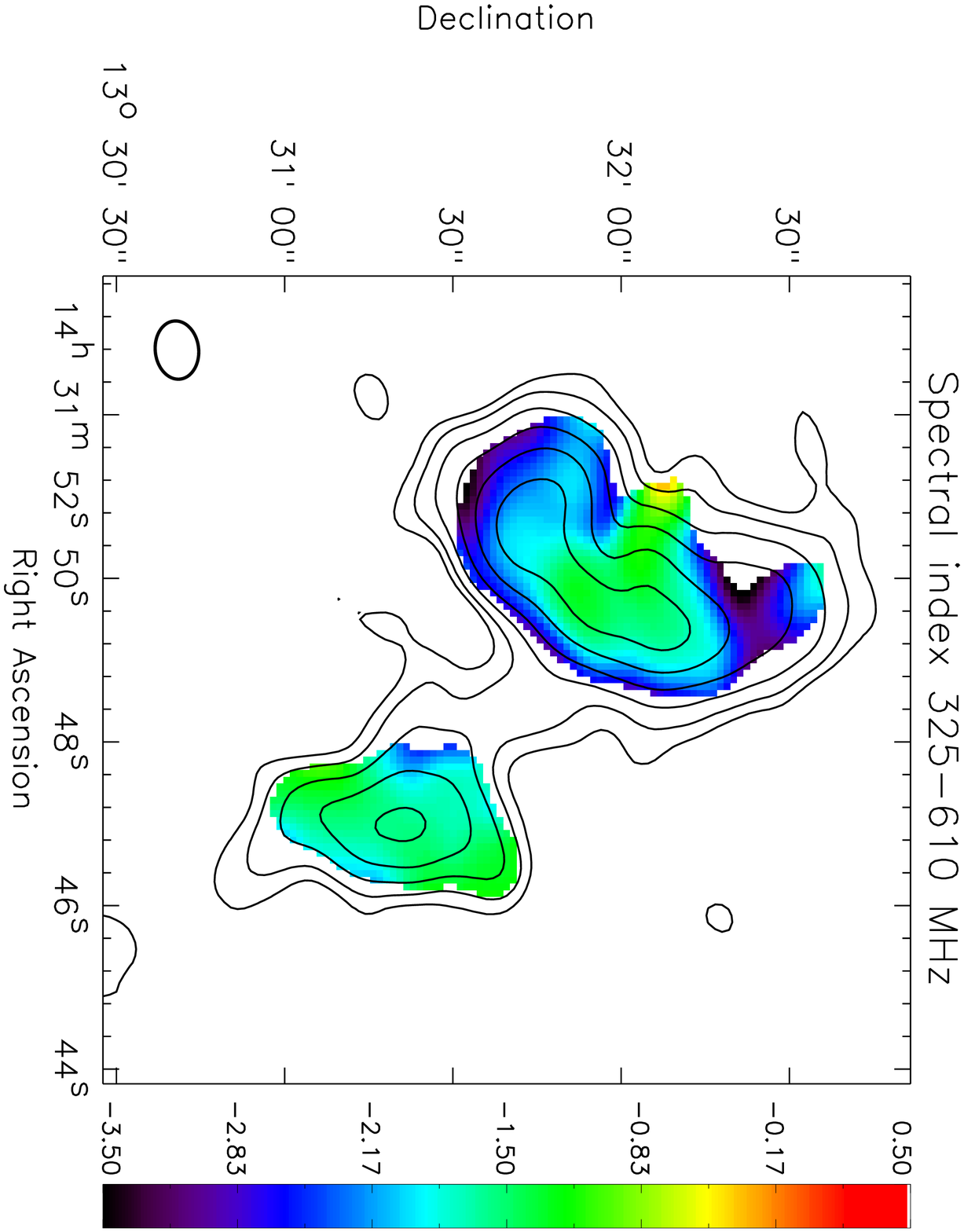}
\includegraphics[angle =90, trim =0cm 0cm 0cm 0cm,width=0.32\textwidth]{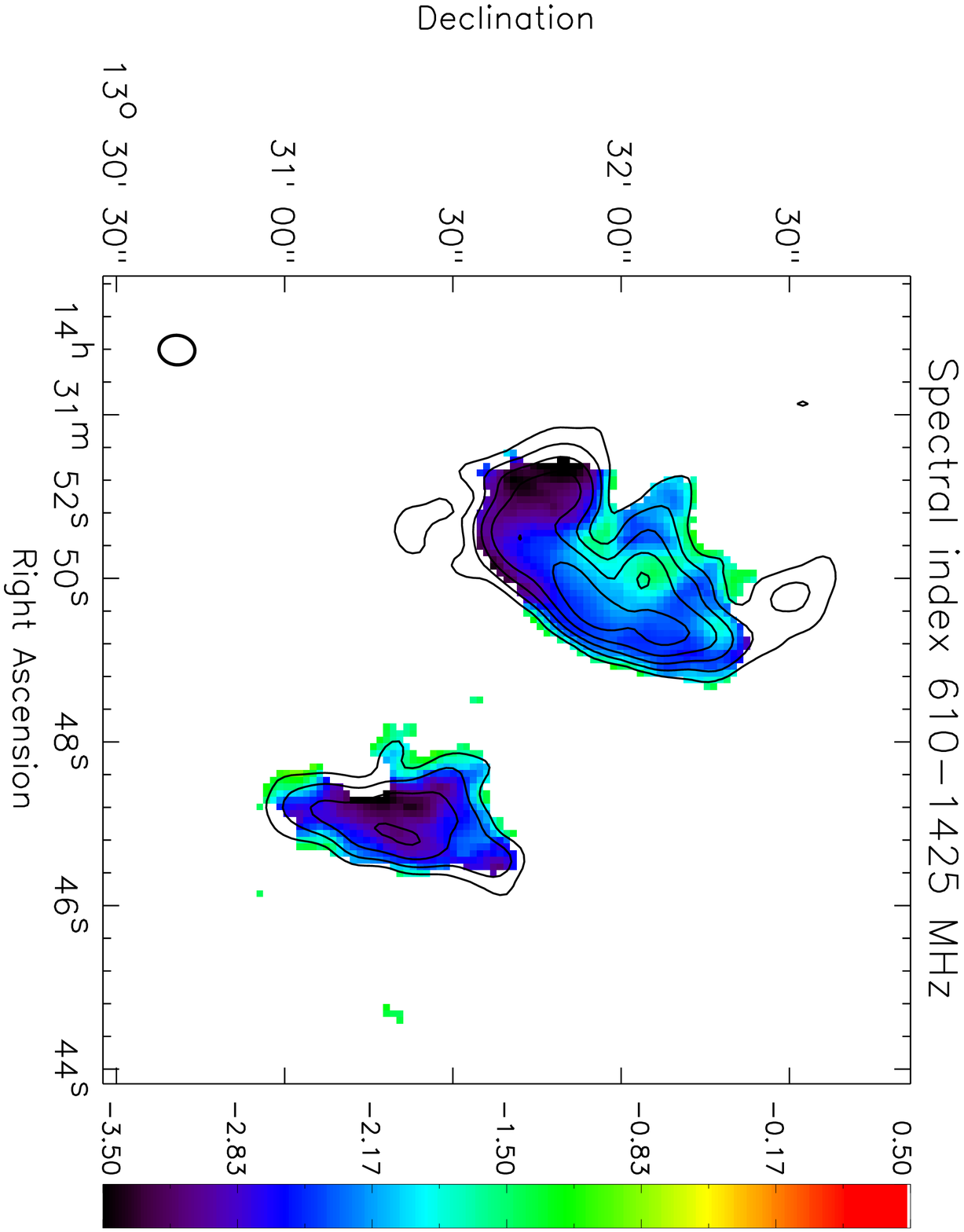}
\includegraphics[angle =90, trim =0cm 0cm 0cm 0cm,width=0.32\textwidth]{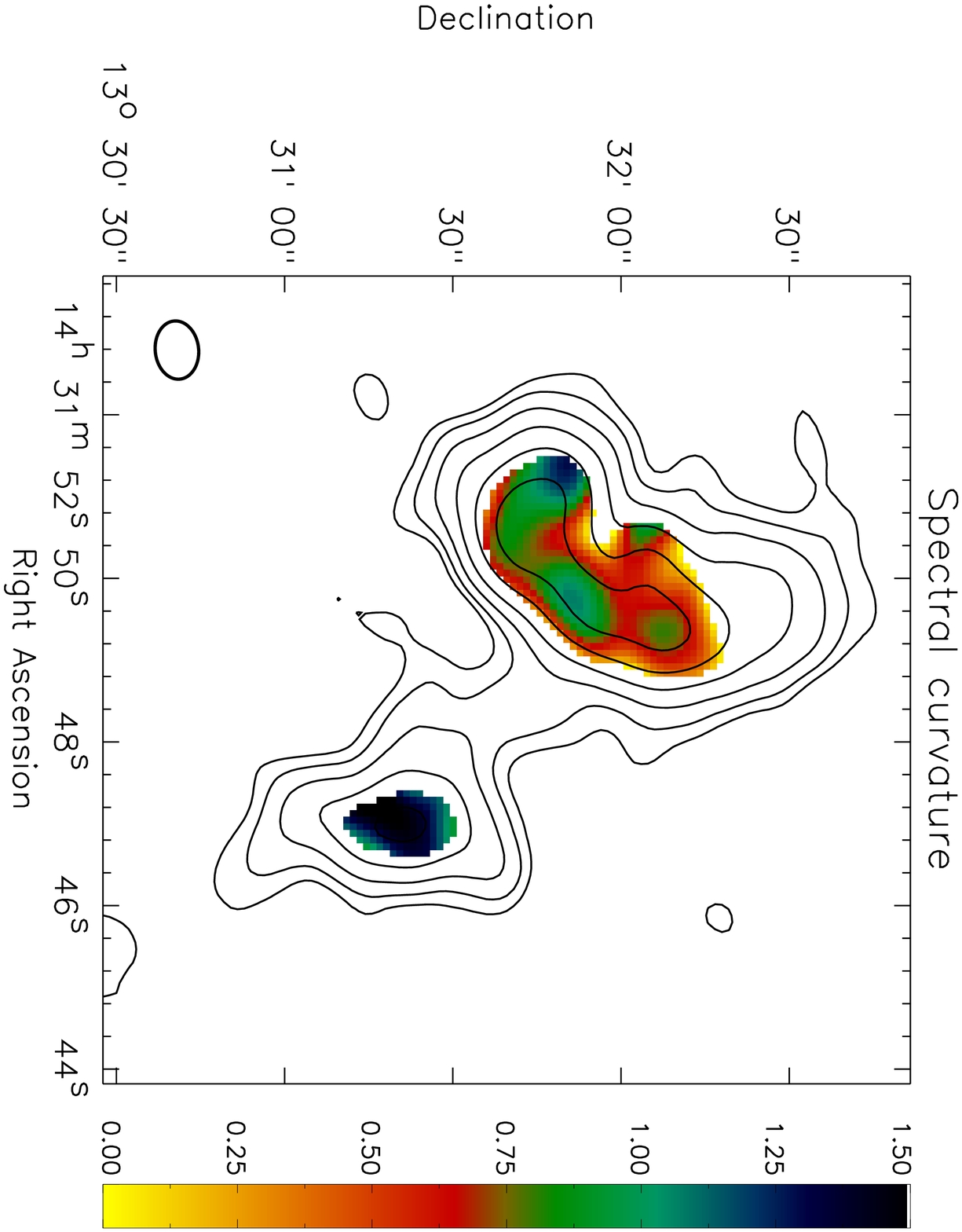}
\end{center}
\caption{Top left: GMRT 325 MHz map. Contour levels are drawn as in Fig.~\ref{fig:gmrt325_5}. 
Top middle: VLA 1425~MHz map. Contour levels are drawn as in Fig.~\ref{fig:gmrt325_5}. 
Top right: VLA 1425~MHz high-resolution image. The image was made using Briggs weighting with robust set to $-1$. Contour levels are drawn at $\sqrt{[1, 2, 4, 8, \ldots]} \times 4\sigma_{\mathrm{rms}}$. 
Bottom left: Spectral index map between 325 and 610 MHz at a resolution of $10.4\arcsec\times7.7\arcsec$. Contour levels are from the 325 MHz GMRT image  and drawn at levels of $[1, 2, 4, 8,  \ldots]~\times 6\sigma_{\mathrm{rms}}$. Bottom middle: Spectral index between 610 and 1425~MHz.  Contour levels are from the 610~MHz GMRT image and drawn at levels of $[1, 2, 4, 8,  \ldots]~\times 5\sigma_{\mathrm{rms}}$. The beam size is $6.41\arcsec \times 5.26\arcsec$. Bottom right: Spectral curvature map.  Contour levels are drawn as in the left panel and the resolution is $10.4\arcsec \times 7.7\arcsec$.}
\label{fig:gmrt325_s6}
\end{figure*}

\begin{figure}
\begin{center}
\includegraphics[angle =90, trim =0cm 0cm 0cm 0cm,width=0.5\textwidth]{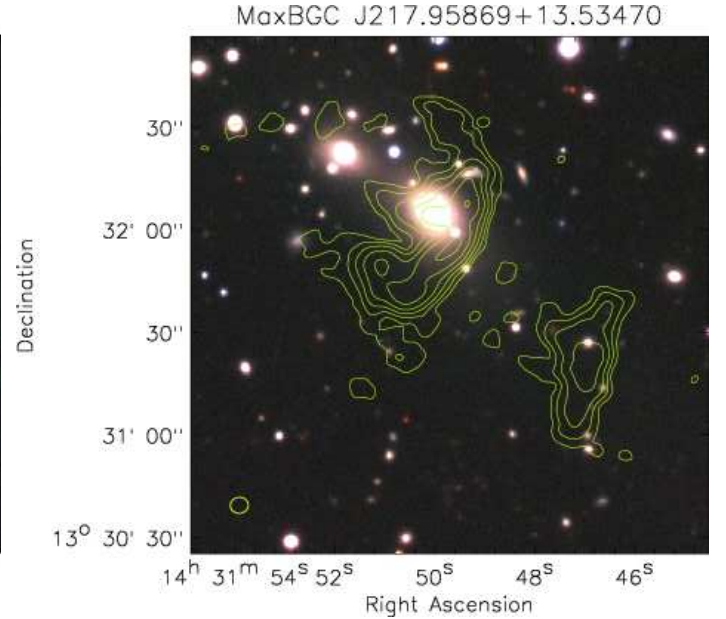}
\end{center}
\caption{Optical WHT color image for MaxBCG~J217.95869+13.53470. GMRT 610~MHz contours are overlaid in yellow. The beam size is $5.3\arcsec \times 4.8\arcsec$. Contour levels are drawn at $[1, 2, 4, 8, \ldots] \times 4\sigma_{\mathrm{rms}}$.} 
\label{fig:s6_optical}
\end{figure}

VLSS~J1431.8+1331 is located in the galaxy cluster 
\object{MaxBCG~J217.95869+13.53470} \citep[$z= 0.1599$,][]{2007ApJ...660..239K} 
and associated with the  central cD galaxy of the cluster. 
The cluster has a moderate X-ray luminosity of  $L_{\rm{X,~0.1-2.4~keV}} \sim 1 \times 10^{44}$~erg~s$^{-1}$ based 
on the ROSAT count rate \citep{1999A&A...349..389V}.  
The GMRT 325~MHz image (see top left panel Fig.~\ref{fig:gmrt325_s6}) 
shows a bright elongated source. To the west a somewhat 
fainter diffuse component can be seen. This component is not associated with any 
optical galaxy (see Fig.~\ref{fig:s6_optical}). A third 
fainter source is located further to the southwest. The first two 
components are connected by a faint radio ``bridge''. 
This bridge was not seen in our previous 610~MHz image. 
The bright source is a currently active radio galaxy with 
the radio core clearly being visible in our VLA 1.4~GHz images  
(see Fig.~\ref{fig:gmrt325_s6} top middle and right panels). 
Probably, radio plasma from the core flows westwards and 
then forms the north-south elongated structure.

The spectral index maps (see Fig.~\ref{fig:gmrt325_s6} bottom panels), are indicative of a 
relatively flat spectral index of $-0.5$ for the radio core 
between 325 and 610~MHz. Spectral steepening is observed to 
the north and south of the elongated structure. The spectral 
index for the southern part of the elongated structure steepens 
to $-3.5$ between 610 and 1425~MHz.  The spectral index of the 
southwestern component is about $-1.5$ between 325 and 610~MHz, 
there being smaller spectral index variations across it than in the 
brighter western component. Between 610 and 1425~MHz, the spectral 
index steepens to about $-2.5$. The spectral curvature 
map ($\alpha_{325-610} - \alpha_{610-1425}$) (see Fig.~\ref{fig:gmrt325_s6} 
bottom right panel), shows that the southwestern source has a very curved 
radio spectrum. The southern end of the radio structure from 
the active AGN is also quite curved. The high spectral curvature 
is likely to be the result of spectral ageing, the gradient in the 
spectral index away from the core providing evidence of this. 
The two diffuse sources the southwest of the active AGN are probably 
old ``bubbles'' of radio plasma linked to this AGN, which is consistent with 
the curved radio spectrum. The presence of a faint radio bridge also 
suggests a relation between this southwestern component and the radio 
galaxy.  The southwestern component can therefore be classified as a 
radio phoenix (if the radio plasma has been compressed) or an AGN relic. 
XMM-Newton observations of the cluster will be presented by Ogrean et al. (submitted).

\subsection{VLSS J2217.5+5943, 24P73} 
\begin{figure*}
\begin{center}
\includegraphics[angle =90, trim =0cm 0cm 0cm 0cm,width=0.45\textwidth]{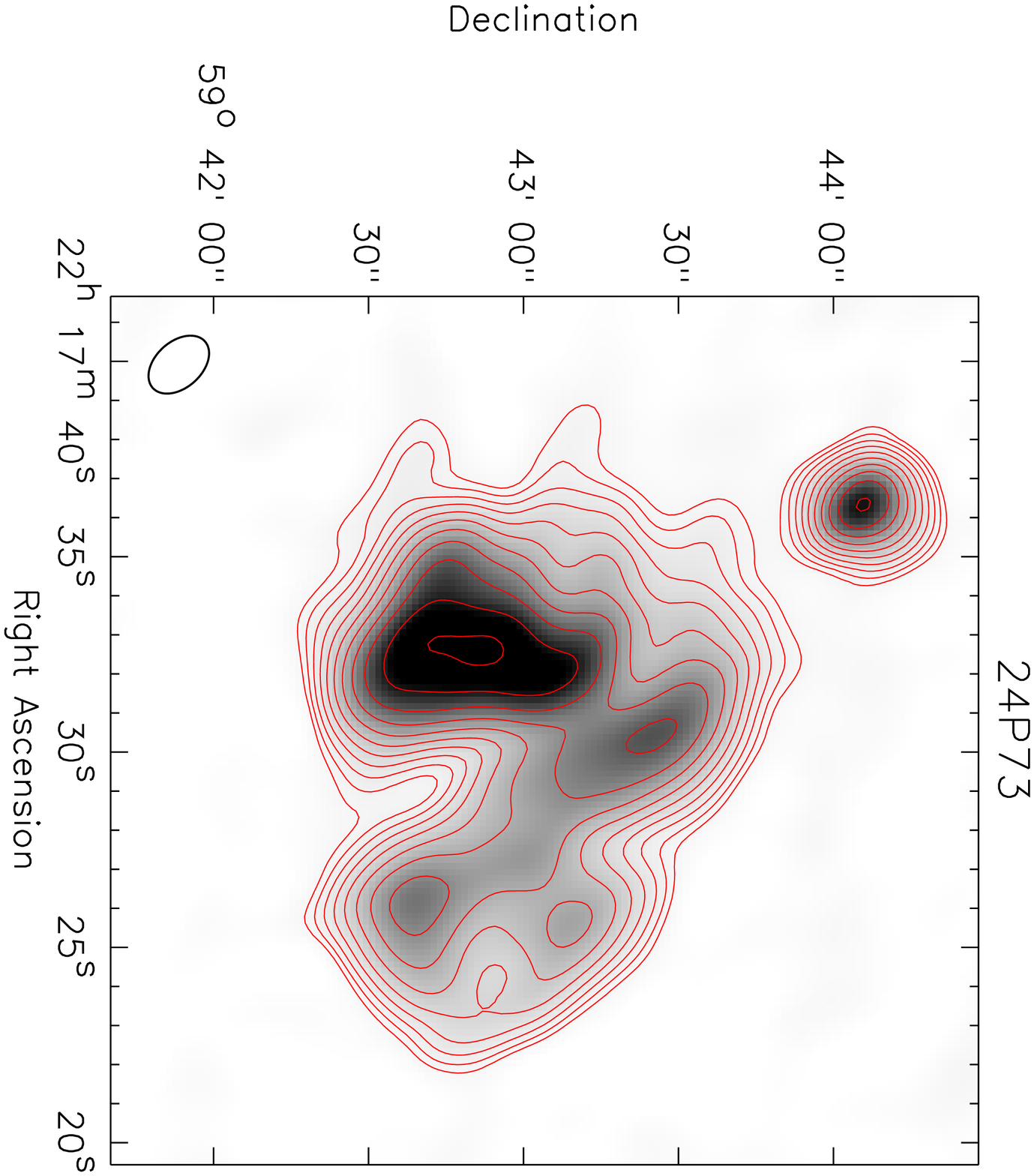}
\includegraphics[angle =90, trim =0cm 0cm 0cm 0cm,width=0.45\textwidth]{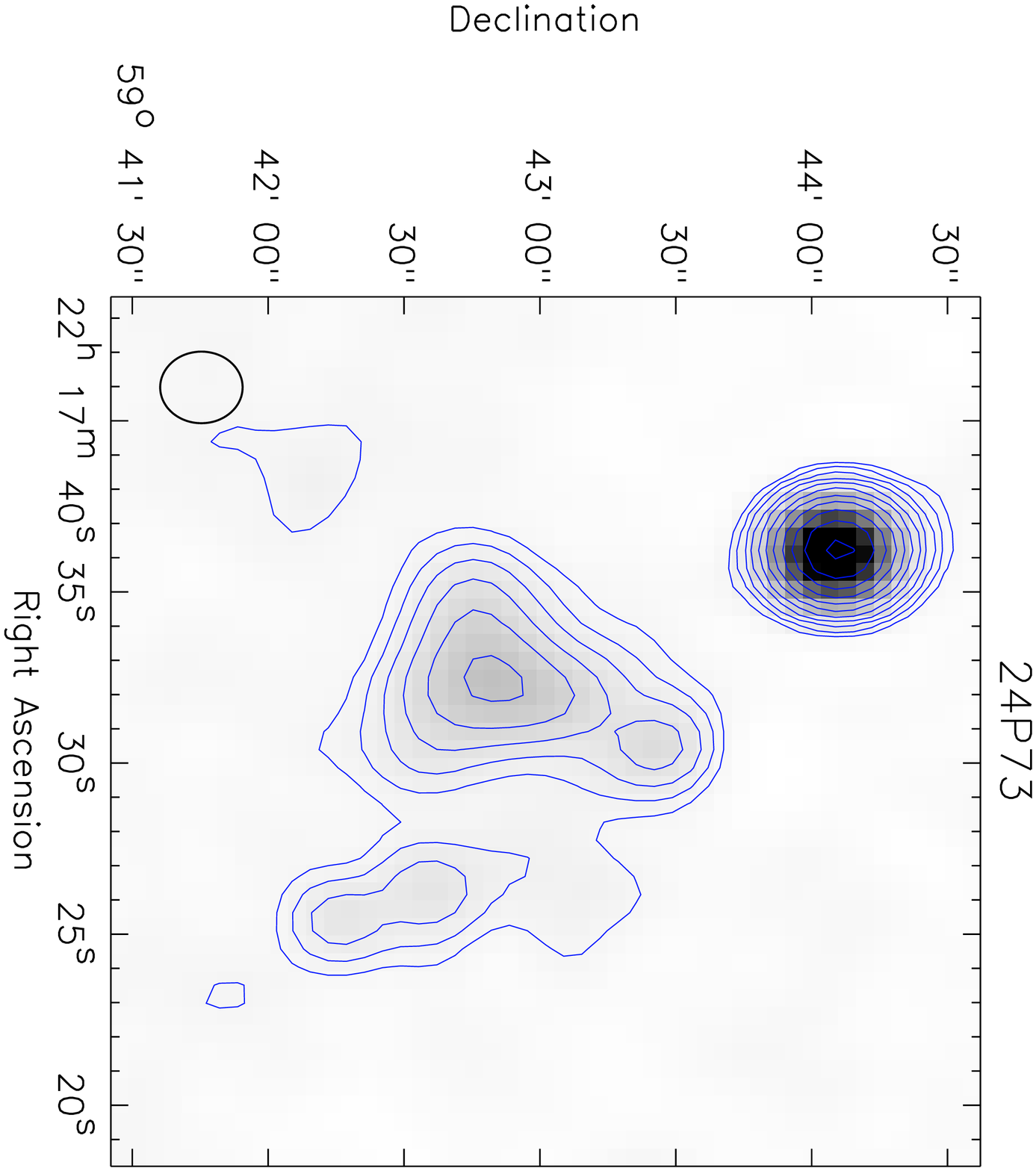}
\includegraphics[angle =90, trim =0cm 0cm 0cm 0cm,width=0.45\textwidth]{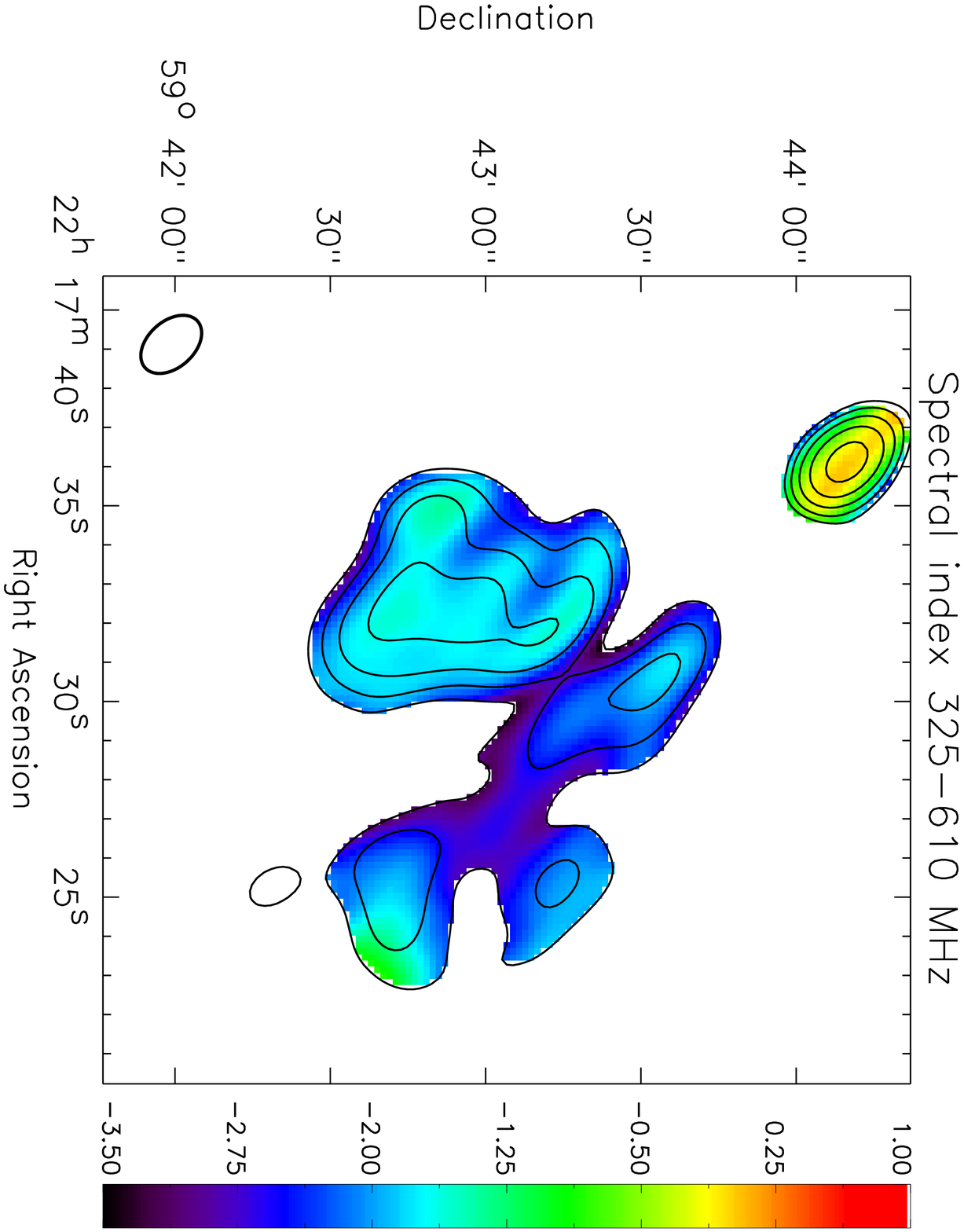}
\includegraphics[angle =90, trim =0cm 0cm 0cm 0cm,width=0.45\textwidth]{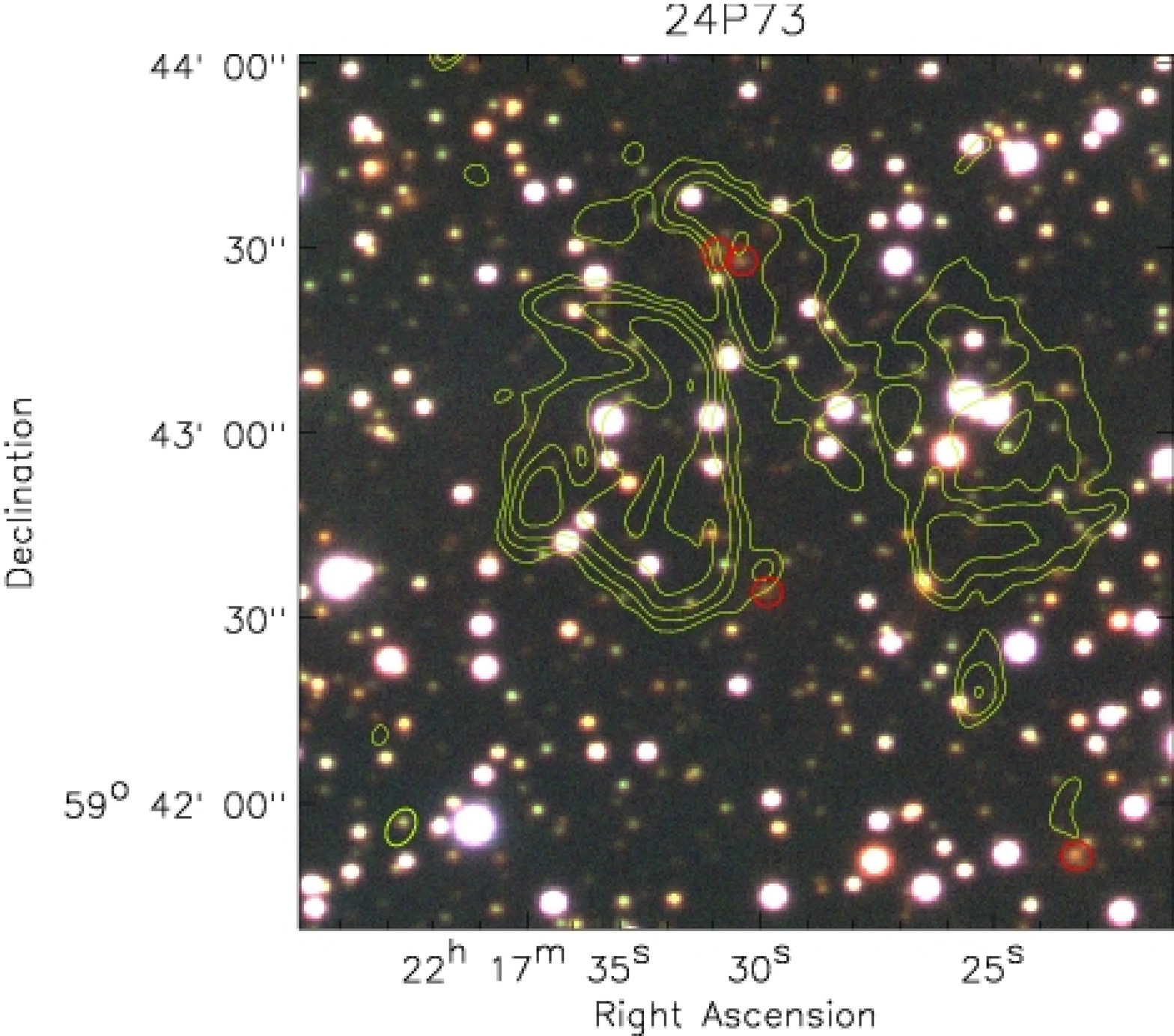}
\end{center}
\caption{Top left: GMRT 325 MHz map. Contour levels are drawn as in Fig.~\ref{fig:gmrt325_5}. Top right: WSRT $1.3-1.8$ GHz map. Contour levels are drawn as in Fig.~\ref{fig:gmrt325_5}. Bottom left: Spectral index map between 325 and 610~MHz at a resolution of $13.44\arcsec\times8.96\arcsec$. Contour levels are from the 325~MHz image and drawn as in Fig.~\ref{fig:spix5}. Bottom right: Optical WHT color image for 24P73. GMRT 610~MHz contours are overlaid in yellow. The beam size is $6.4\arcsec \times 4.3\arcsec$. Contour levels are drawn as in Fig.\ref{fig:s6_optical}. Several faint galaxies in the image are marked by red circles.}
\label{fig:gmrt325_14}
\end{figure*}

This source was discovered during the Synthesis Telescope of the Dominion Radio Observatory (DRAO) Galactic plane survey at 408~MHz and 1.42~GHz \citep{1989JRASC..83..105H,1990A&AS...82..113J}.  \cite{1994A&AS..104..481G} found the source to be diffuse and have an ultra-steep spectrum ($\alpha=-2.58 \pm 0.14$). Our GMRT 610~MHz observations (see Fig.~\ref{fig:gmrt325_14} bottom right panel) detected a very complex filamentary source, resembling the relics found in \object{Abell~13} and \object{Abell~85} \citep{2001AJ....122.1172S}. Our GMRT 325~MHz image, Fig.~\ref{fig:gmrt325_14} top left panel, is similar to the 610~MHz image. In our combined WSRT $1.3-1.8$~GHz image (top right panel),  the fainter western part of the relic is only marginally detected,  which is indicative of a steep spectral index in  this region. We set an upper limit to the polarization fraction of 5\% for the source at 1.4~GHz.

The spectral index map between 325 and 610~MHz is shown in the bottom left panel of Fig.~\ref{fig:gmrt325_14}. The spectral index varies between $-1.4$ and $-2.9$ over the source. The western part of the source has the steepest spectrum. Towards the southwest, the spectral index flattens to about $-1.0$. This part may be associated with a separate compact radio AGN.

Our optical WHT image (bottom right panel Fig.~\ref{fig:gmrt325_14}) is dominated by foreground stars as the source is located in the Galactic plane, at Galactic latitude of  $2.44$ degrees. However, there are also several faint red galaxies seen in the image that may belong to a galaxy cluster, and these are marked by red circles. The brightest galaxy (located southwest) has an R-band magnitude of $20.69$. Using the Hubble-R relation \citep{2007A&A...464..879D}, we estimate a redshift of $0.15\pm0.1$, including an extinction in the R-band of 4.173 mag \citep{1998ApJ...500..525S}. We note that this redshift estimate is based on the corrected identification of the cD galaxy in the cluster. If the source is indeed located at $z=0.15$, then its largest physical extent is 270~kpc. We classify the source as a radio phoenix given the filamentary morphology and extreme spectral index. In fact, the relic is very similar to the proposed phoenix in \object{Abell 13} \citep{2001AJ....122.1172S}. To confirm the presence of a cluster, deep near-infrared (NIR) imaging will be necessary.

\subsection{VLSS J0004.9$-$3457} 
\begin{figure*}
\begin{center}
\includegraphics[angle =90, trim =0cm 0cm 0cm 0cm,width=0.45\textwidth]{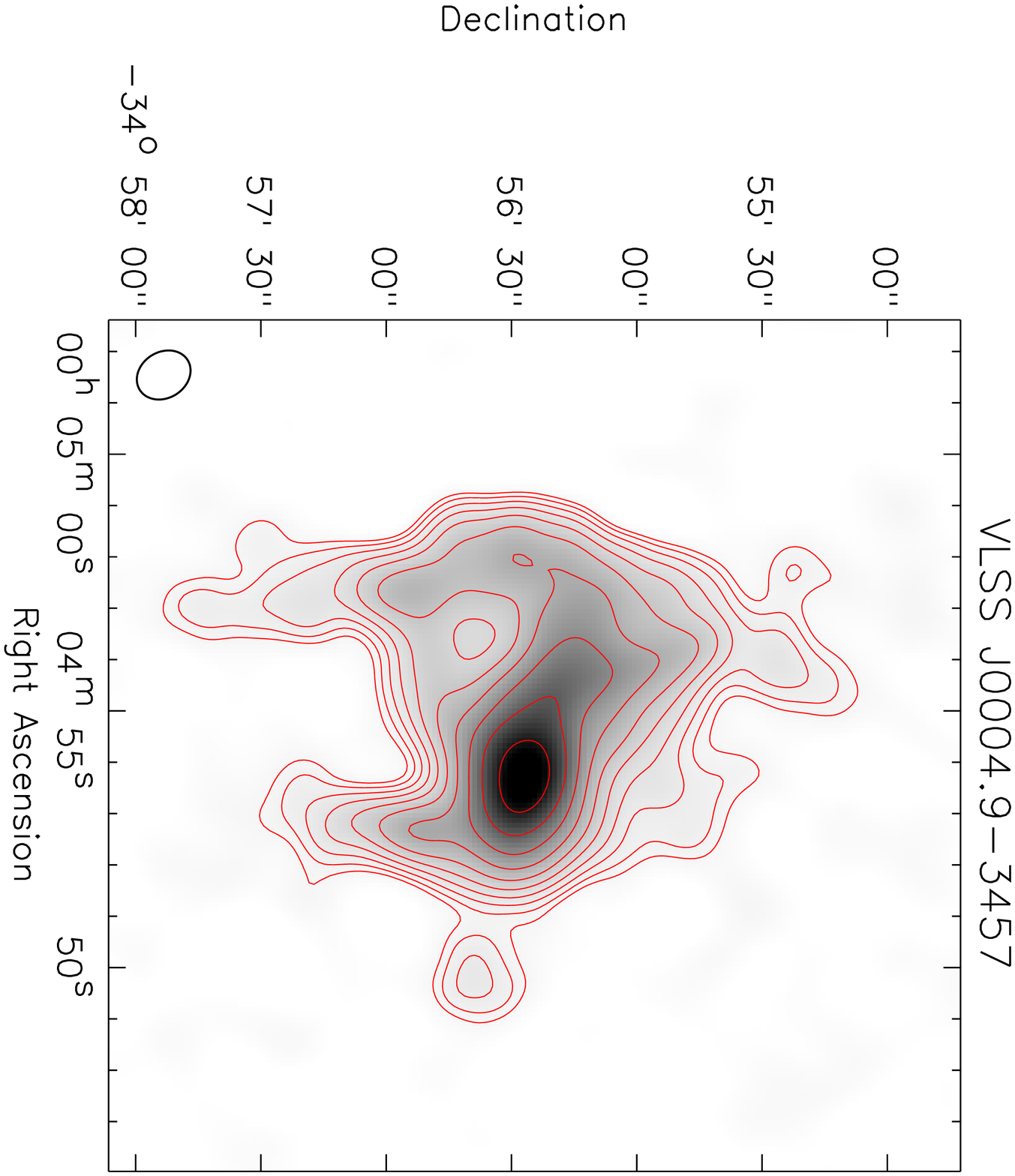}
\includegraphics[angle =90, trim =0cm 0cm 0cm 0cm,width=0.45\textwidth]{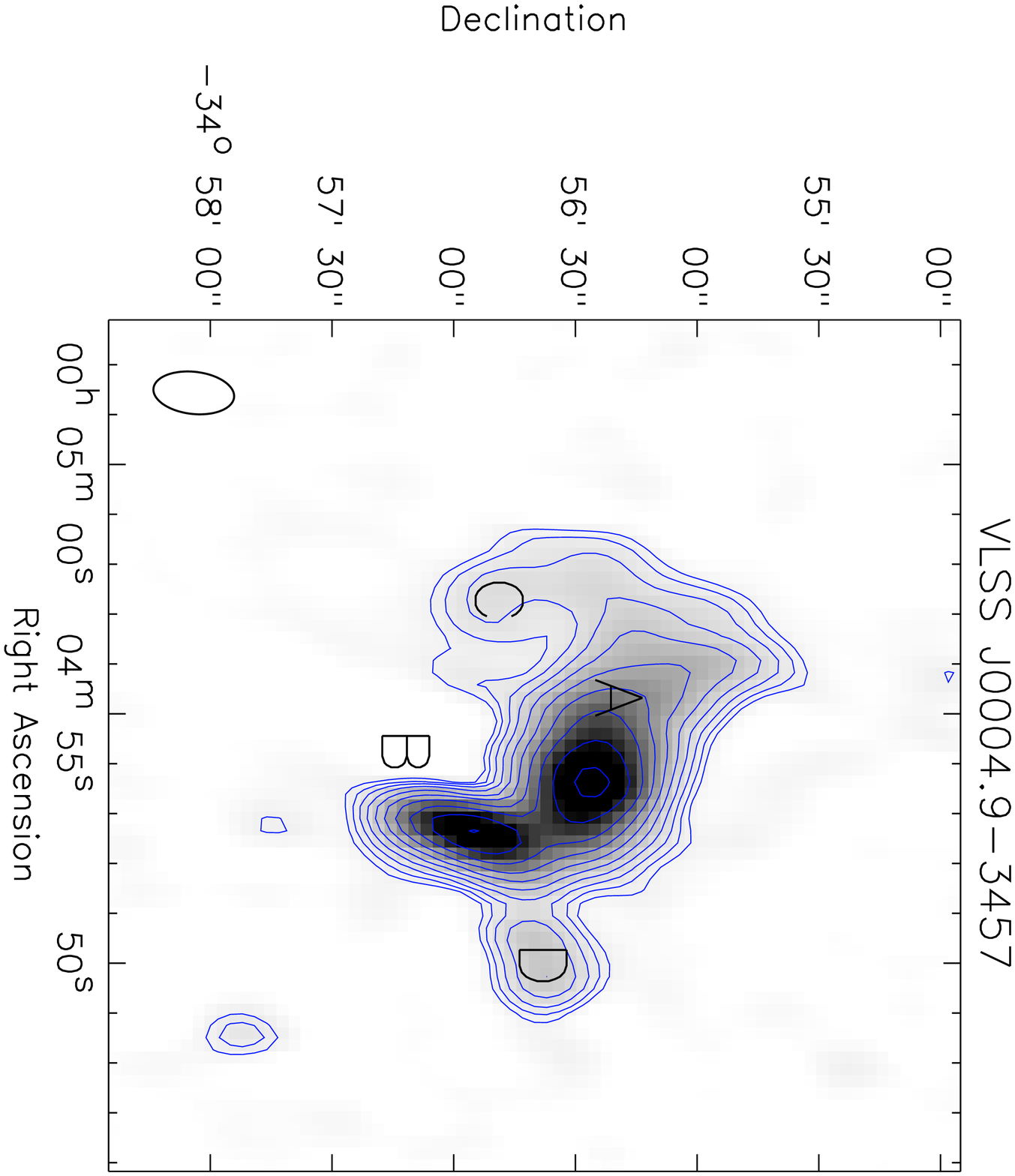}
\includegraphics[angle =90, trim =0cm 0cm 0cm 0cm,width=0.45\textwidth]{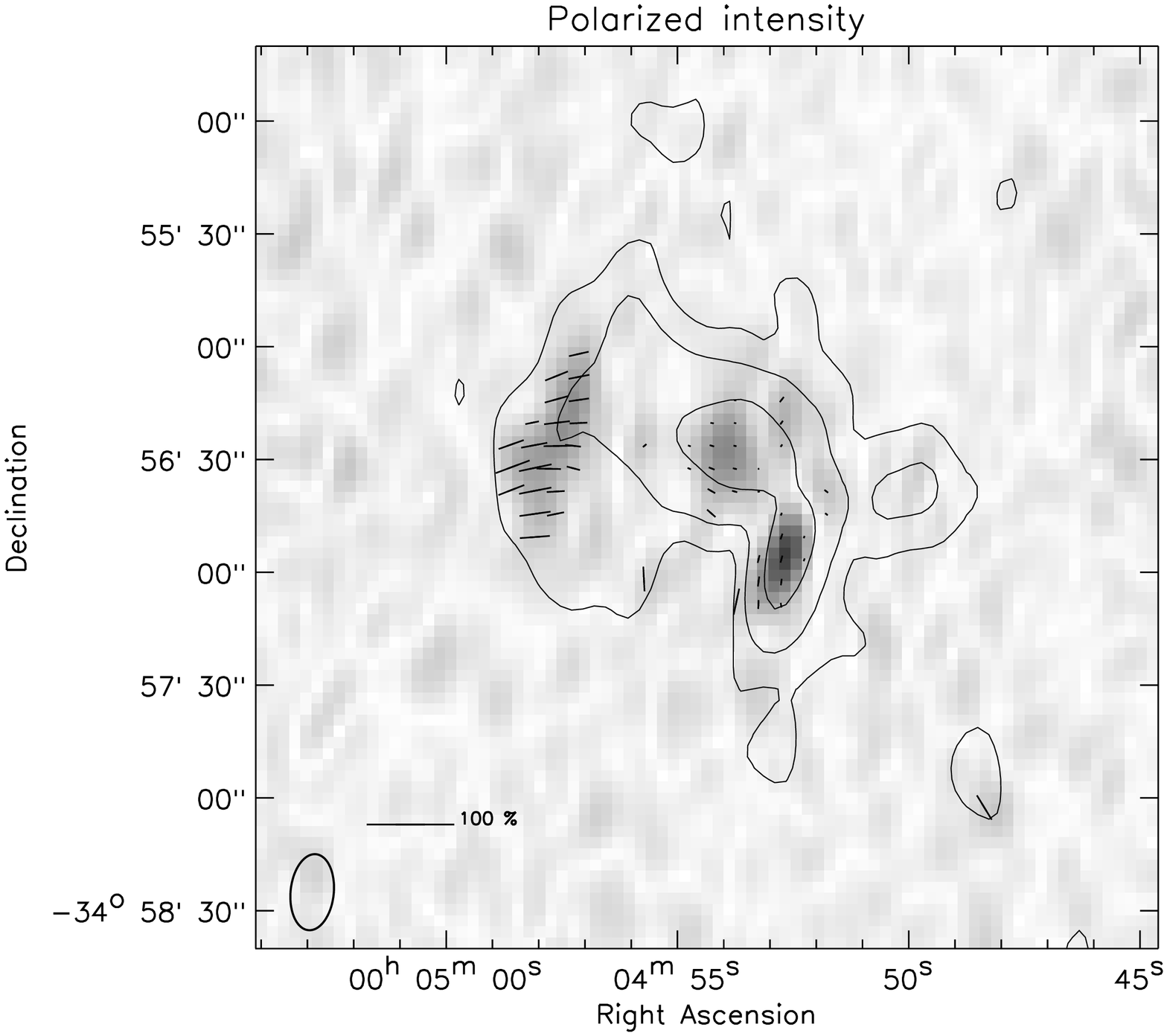}
\includegraphics[angle =90, trim =0cm 0cm 0cm 0cm,width=0.45\textwidth]{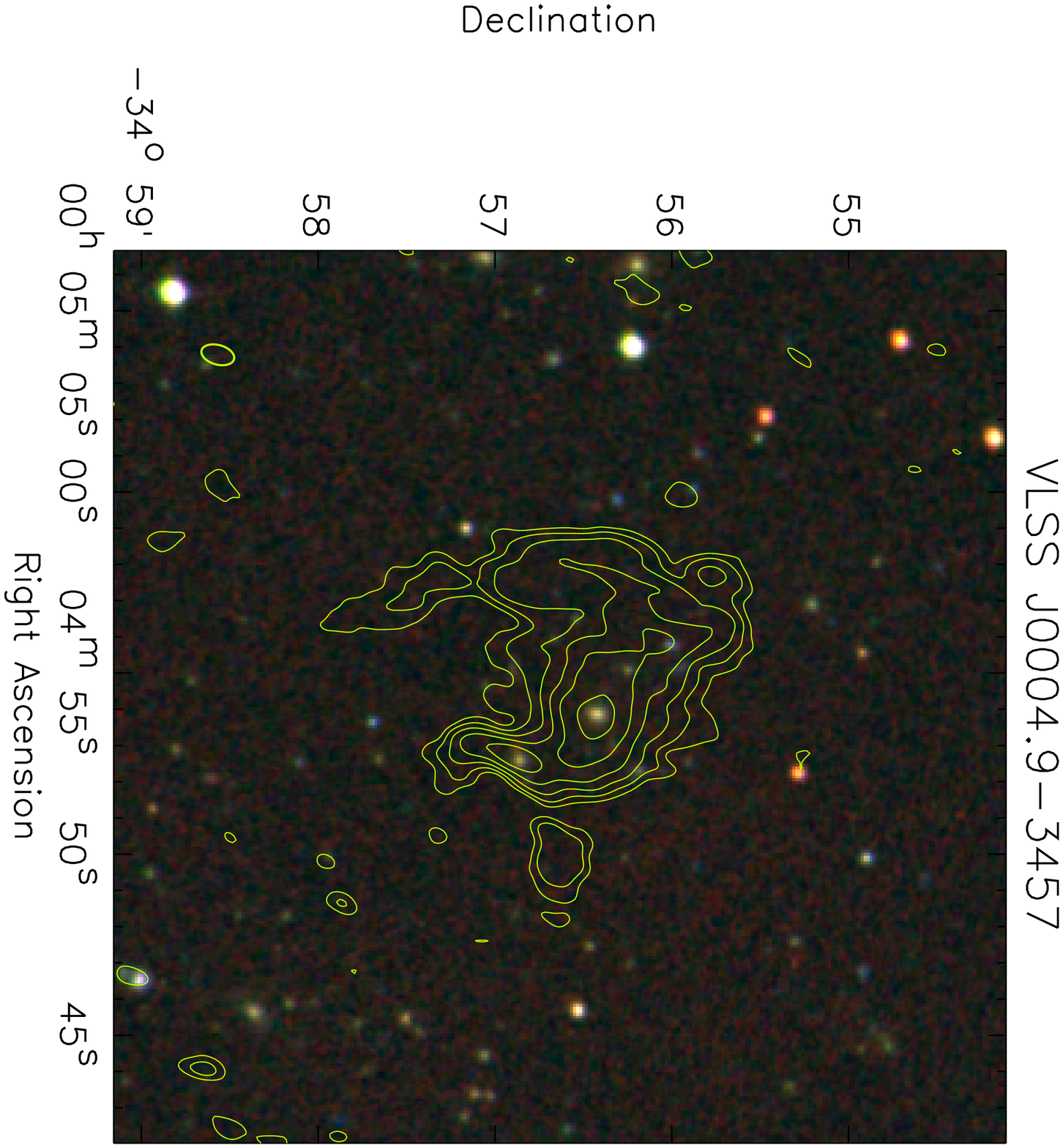}
\end{center}
\caption{Top left: GMRT 325~MHz map. Contour levels are drawn as in Fig.~\ref{fig:gmrt325_5}.  
Top right: VLA 1425~MHz map. Contour levels are drawn as 
in Fig.~\ref{fig:gmrt325_5}.  
Bottom left: VLA 1425~MHz polarization map. Total polarized intensity is shown 
as grayscale image. 
Vectors refer to the polarization E-vectors, with their length representing the 
polarization fraction. A reference vector for a polarization fraction of 
100\% is shown in the bottom left corner. The polarization fractions were corrected for Ricean bias \citep{1974ApJ...194..249W}. No polarization E-vectors were drawn for pixels with a SNR less than 3 in the total polarized intensity map. Contours show the total intensity image (Stokes I) at 1425~MHz. 
Contour levels are drawn at $[1, 16, 256, 4096, \ldots]~\times 0.147$~mJy~beam$^{-1}$. 
Bottom right: Optical POSS-II color image for VLSS~J0004.9$-$3457. 
GMRT 610~MHz contours are overlaid in yellow. The beam size 
is $6.4\arcsec \times 4.3\arcsec$. Contour levels are drawn 
as in Fig.\ref{fig:s6_optical}.}
\label{fig:gmrt325_s18}
\end{figure*}

The radio source is located in a small galaxy cluster or 
group, \object{B02291} \citep{2001A&A...379...21Z}. The cluster/group is located at a redshift of  $0.3 \pm 0.1$ 
\citep{2009A&A...508...75V}. No X-ray emission from the 
system is detected in the ROSAT All-Sky Survey 
\citep{1999A&A...349..389V, 2000IAUC.7432R...1V}, 
which implies that the system is not very massive. An 
optical POSS-II color composite is shown in 
Fig.~\ref{fig:gmrt325_s18} (bottom right panel). 
The various radio components are labeled alphabetically  
(see Fig.~\ref{fig:gmrt325_s18} top right panel).

Our GMRT 325~MHz image of \object{VLSS J0004.9$-$3457} 
(top left panel) displays a diffuse source (A) centered on 
a K-mag=14.86 galaxy.  The source extends somewhat 
further northwards than in the 610~MHz image (bottom right panel).

Source B is associated with another galaxy, C does 
not have an optical counterpart and seems to be 
related to source A. Source D is a fainter source 
(resolved in the 610~MHz image) located just east 
of  \object{VLSS J0004.9$-$3457} at 
RA~00$^\mathrm{h}$~04$^\mathrm{m}$~50$^\mathrm{s}$, 
Dec~$-$34\degr~56\arcmin~38\arcsec. In the CnB-array 
VLA 1.4~GHz image, shown in Fig.~\ref{fig:gmrt325_s18} 
(top right), component C is less prominent than in the 
325~MHz image, while source B is clearly visible.

\begin{figure*}
\begin{center}
\includegraphics[angle =90, trim =0cm 0cm 0cm 0cm,width=0.32\textwidth]{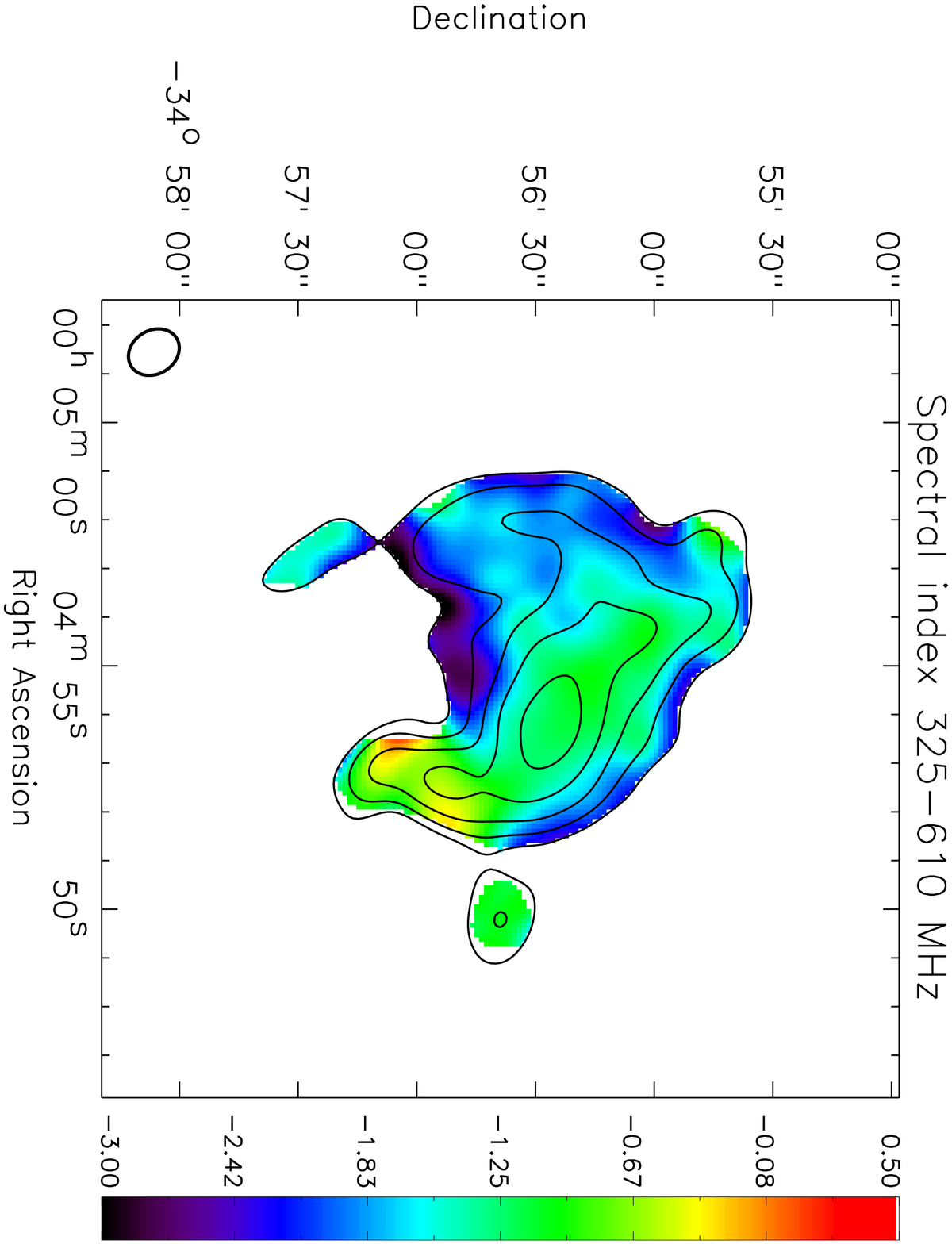}
\includegraphics[angle =90, trim =0cm 0cm 0cm 0cm,width=0.32\textwidth]{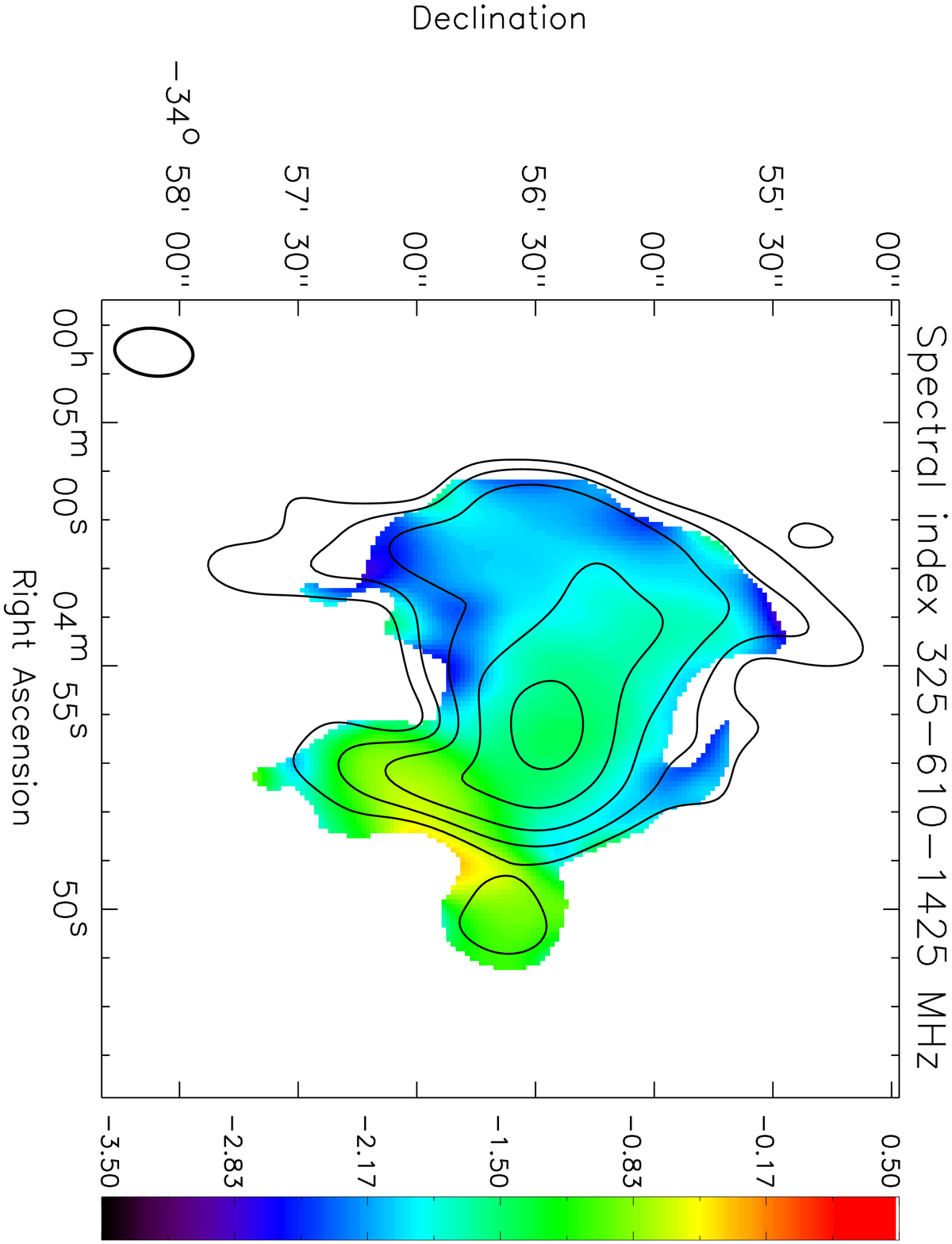}
\includegraphics[angle =90, trim =0cm 0cm 0cm 0cm,width=0.32\textwidth]{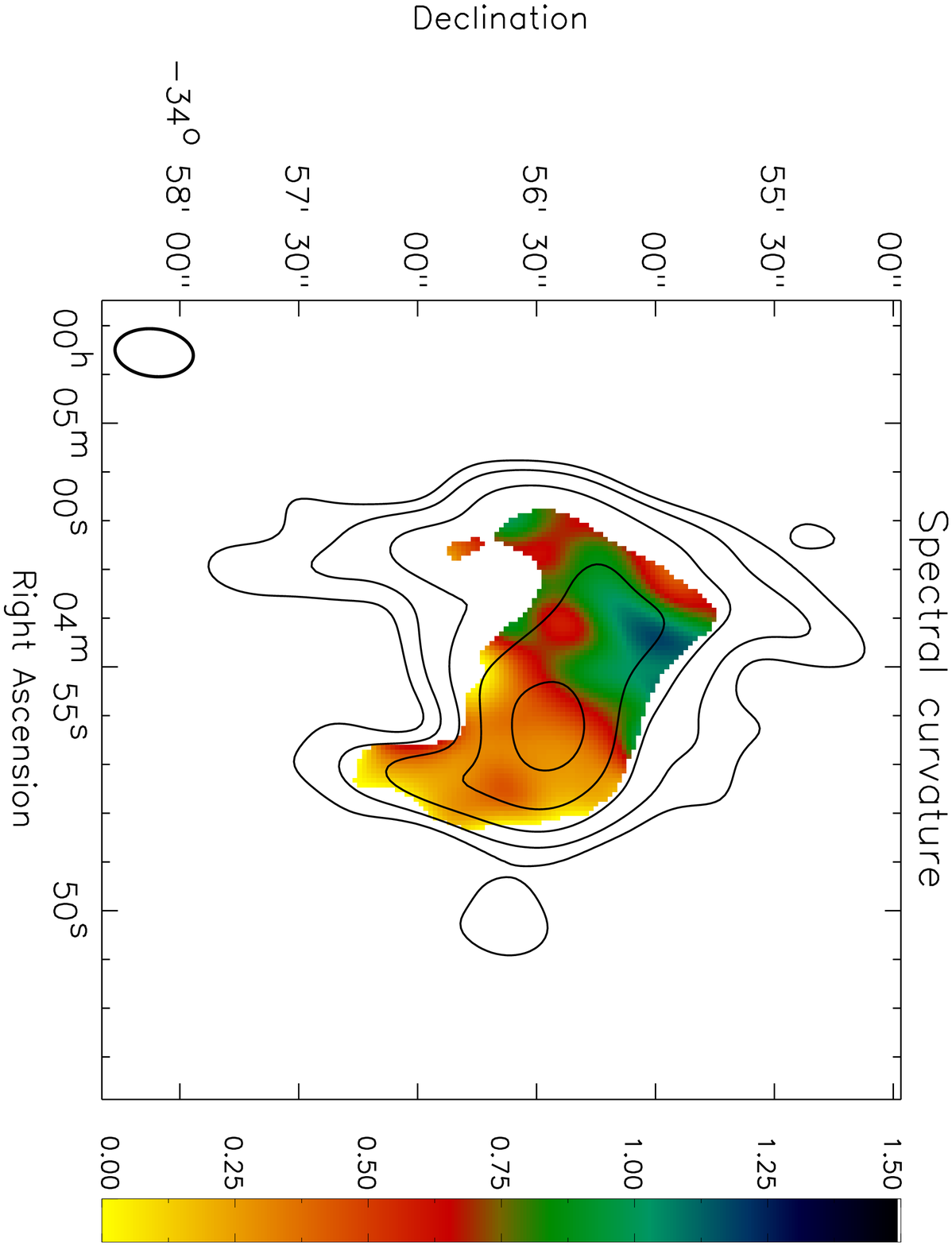}
\end{center}
\caption{Left: Spectral index map between 325 and 610~MHz at a resolution of $13.5\arcsec\times11.0\arcsec$.  Contour levels are from the 325~MHz image and drawn as in Fig.~\ref{fig:spix5}. Middle:  Power-law spectral index fit between 325, 610, and 1425~MHz. Contours are from the 325~MHz image and drawn at levels of $[1, 2, 4, 8, \ldots]~\times 4\sigma_{\mathrm{rms}}$. The beam size is $19.83\arcsec \times 12.0\arcsec$. Right: Spectral curvature map.  Contour are drawn as in the middle panel. }
\label{fig:spix18}
\end{figure*}

The spectral index map between 325 and 610~MHz 
is shown in Fig.~\ref{fig:spix18} (right panel). 
Source B has a flat spectral index of $\sim-0.5$. 
Source D has a steeper spectral index of $-1.1$. 
The spectral index of A steepens away from the center 
(defined as the peak flux and located at the position 
of the  K-mag=14.86 galaxy). The central region has a 
spectral index of $-1.2$. Outwards, the spectral index 
steepens to $< -2$. Component C has a spectral index of 
about $-1.5$. A spectral index map, for the frequencies 325, 610, 
and 1425~MHz, is shown in Fig.~\ref{fig:spix18} (middle panel). 
The spectral curvature map, in the right panel of 
Fig.~\ref{fig:spix18}, shows the least curvature 
(i.e., less than 0.5 units) for the western part 
of A and source B. Emission to the east of A shows 
more spectral steepening, with a curvature of about 
1.0 units.

Polarized emission from source A, B, and component C  
is observed in the VLA images, Fig.~\ref{fig:gmrt325_s18} 
(bottom left panel). Source B is polarized at the 8\% level. 
The polarization fraction at the center of A is roughly 6\%. 
The arc-like southeast extension (C) is highly polarized, with 
a polarization fraction between 25 and 35\%.

We tentatively classify the source as a 
200~kpc ``mini-halo'' (or core-halo), because 
the radio emission of diffuse source A surrounds 
a central galaxy in a cluster or galaxy group. The spectral 
steepening away from the core is then the result of 
synchrotron and IC losses. The eastern arc-like 
extension could be a radio relic based on its elongated 
nature and high polarization fraction. The high 
polarization fraction is indicative of the presence of ordered 
magnetic fields. Therefore it is likely that this component 
traces a region in the ICM that has been compressed by 
a shock wave. Radio mini-halos usually occur in massive 
relaxed galaxy clusters. The possible presence of a radio 
relic indicates merger activity so it is possible that 
the ``mini-halo'' can also be linked to the merger 
activity of the system. Radio plasma from the central 
AGN might have been re-accelerated or compressed by this 
merger event. The source is somewhat similar to 
\object{MRC~0116+111} studied by \cite{2002IAUS..199..159G} and \cite{2009MNRAS.399..601B}. 
We find that MRC~0116+111 exhibits two bubble-like radio lobes, 
whereas VLSS J0004.9$-$3457 seems to consist of a single 
component. The morphology of the source is more similar 
to the core-halo system in \object{ZwCl~1454.8+2223} and 
candidate core-halo system in \object{Abell~3444} 
\citep{2008A&A...484..327V, 2007A&A...463..937V}. 
The cluster/group would make an interesting target 
for future X-ray observations to study the relation 
between the radio sources and the surrounding ICM.

\subsection{VLSS~J0915.7+2511, MaxBCG~J138.91895+25.19876}  
\begin{figure*}
\begin{center}
\includegraphics[angle =90, trim =0cm 0cm 0cm 0cm,width=0.3\textwidth]{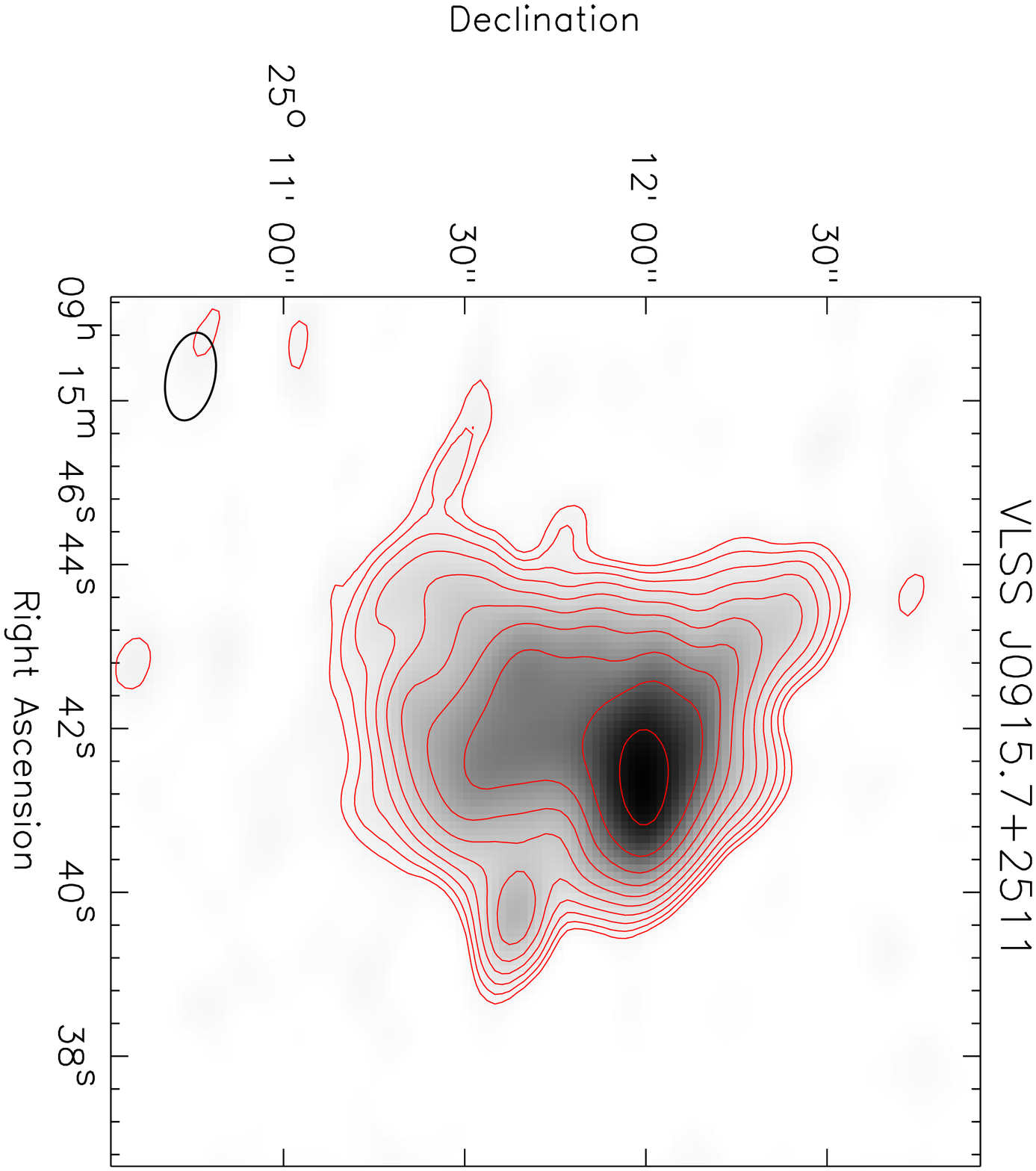}
\includegraphics[angle =90, trim =0cm 0cm 0cm 0cm,width=0.35\textwidth]{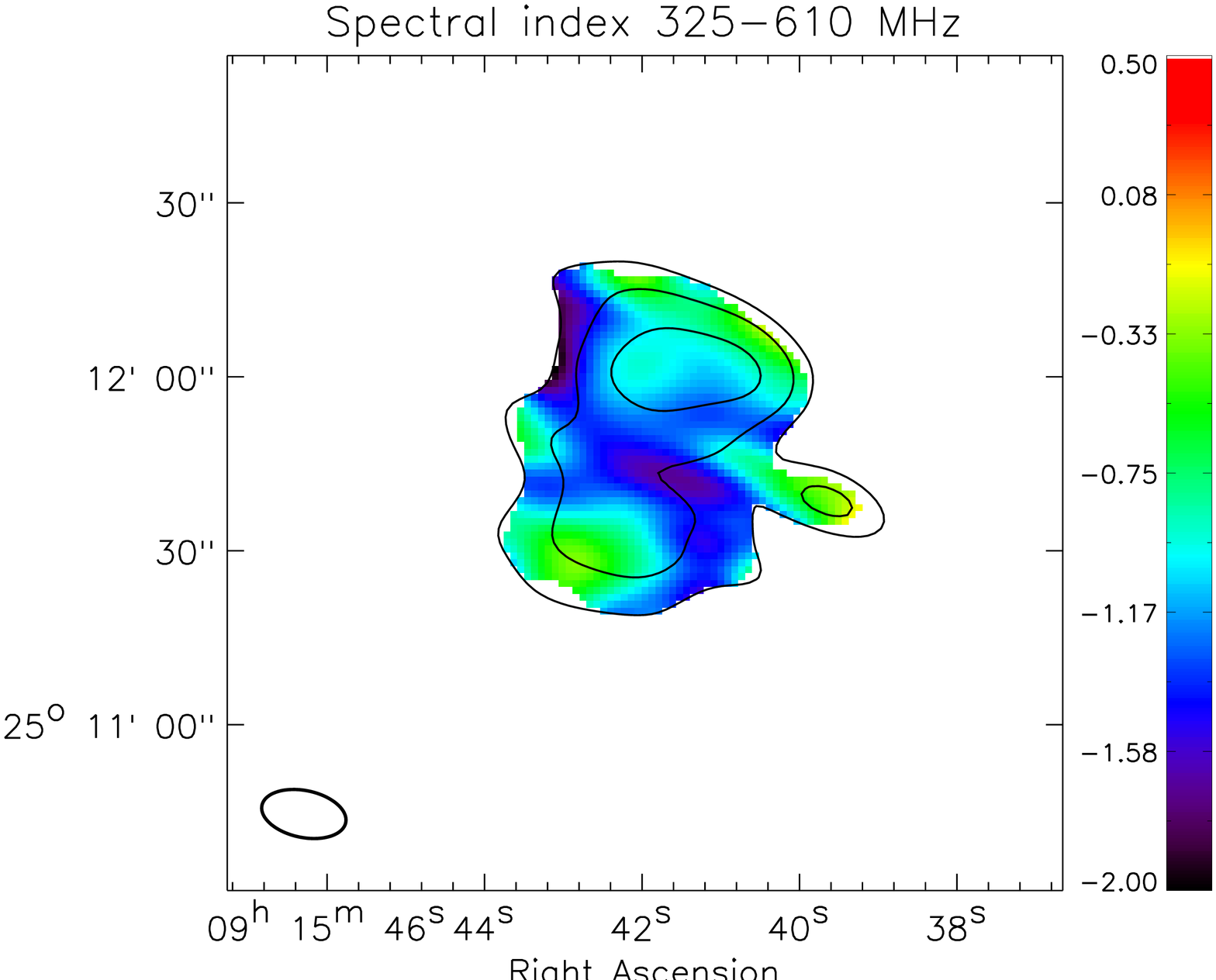}
\includegraphics[angle =90, trim =0cm 0cm 0cm 0cm,width=0.3\textwidth]{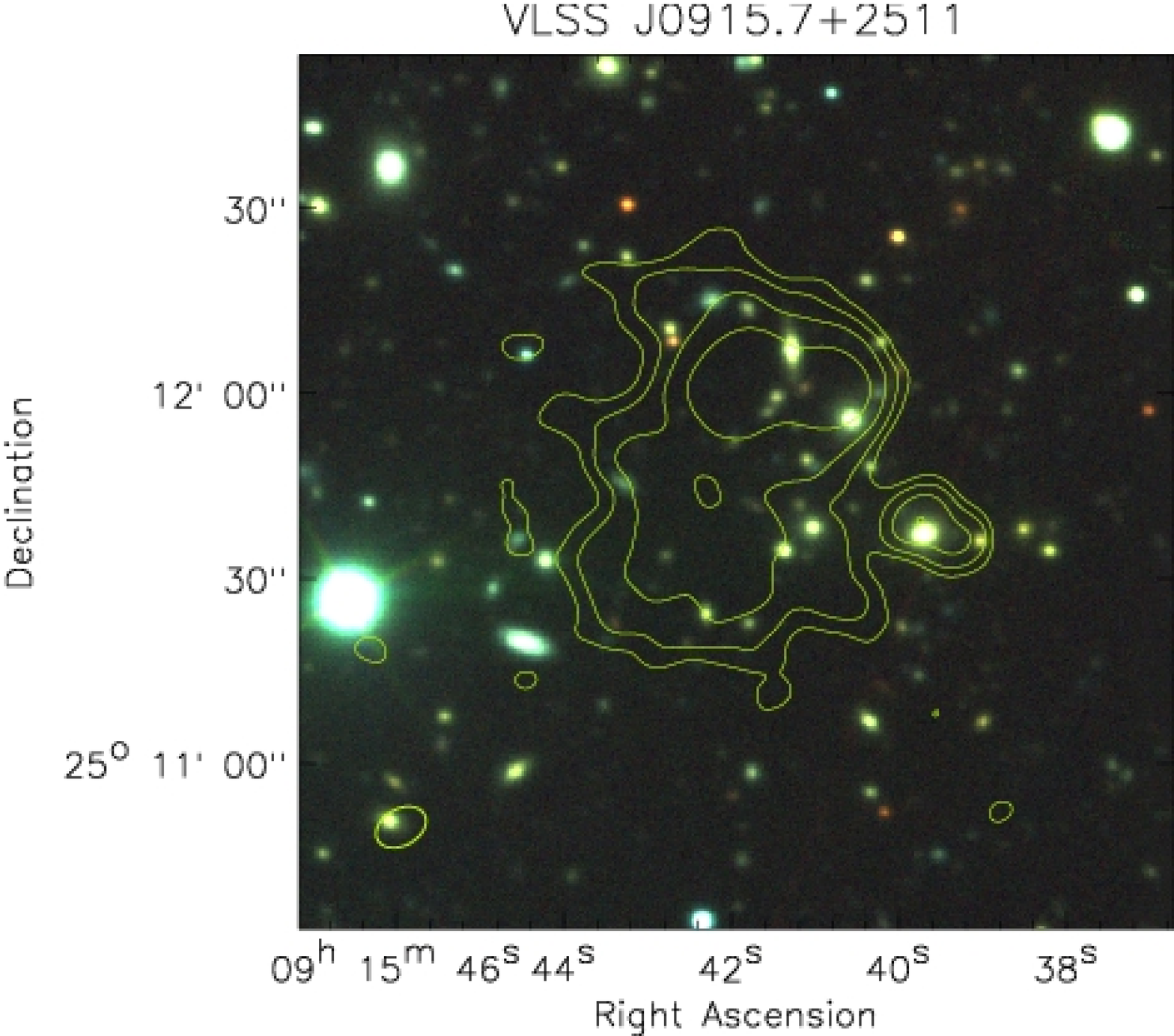}
\end{center}
\caption{Left: GMRT 325 MHz map. Contour levels are drawn as in Fig.~\ref{fig:gmrt325_5}. Middle: Spectral index map between 325 and 610~MHz at a resolution of $14.7\arcsec\times8.1\arcsec$.  Contour levels are from the 325~MHz image and drawn as in Fig.~\ref{fig:spix5}. Right: Optical WHT color image for VLSS~J0915.7+2511. GMRT 610~MHz contours are overlaid in yellow. The beam size is $8.6\arcsec \times 5.9\arcsec$. Contour levels are drawn as in Fig.\ref{fig:s6_optical}.}
\label{fig:gmrt325_22}
\end{figure*}

The radio source is located in the cluster \object{MaxBCG~J138.91895+25.19876}. The source consists of a northern component (Fig.~\ref{fig:gmrt325_22} left panel) and a slightly more extended fainter component to the south. A compact source to the west is associated with the galaxy \object{SDSS~J091539.68+251136.9}. This source has a spectroscopic redshift \citep[SDSS~DR7, ][]{2009ApJS..182..543A}  of 0.324. This cluster has a photometric redshift of $0.289$ \citep{2007ApJ...660..239K}, but the galaxy seems to be part of the cluster, hence we adopt a redshift of $0.324$ for the cluster. The compact source is resolved in our 610~MHz GMRT image and displays a double lobe structure (see Fig.\ref{fig:gmrt325_22} right panel). 

The spectral index map between 325 and 610~MHz is shown in Fig.~\ref{fig:gmrt325_22} (middle panel). The spectral index map is noisy because of dynamic range limitations from the 
source \object{4C~+25.24} (1.35~Jy at 325~MHz) located about 3\arcmin~to the southeast. The eastern part of the northern component has the steepest spectrum with an index of $\alpha \sim -2$, although the SNR is low in this region. The compact source to the west has a flat spectral index of about $-0.2$. 

The classification of the source is unclear. The source might contain old radio plasma that originated in the AGN to the west. In this case, the source could be classified as a radio phoenix or AGN relic. 

\subsection{VLSS~J1515.1+0424, Abell 2048}   
\begin{figure*}
\begin{center}
\includegraphics[angle =90, trim =0cm 0cm 0cm 0cm,width=0.45\textwidth]{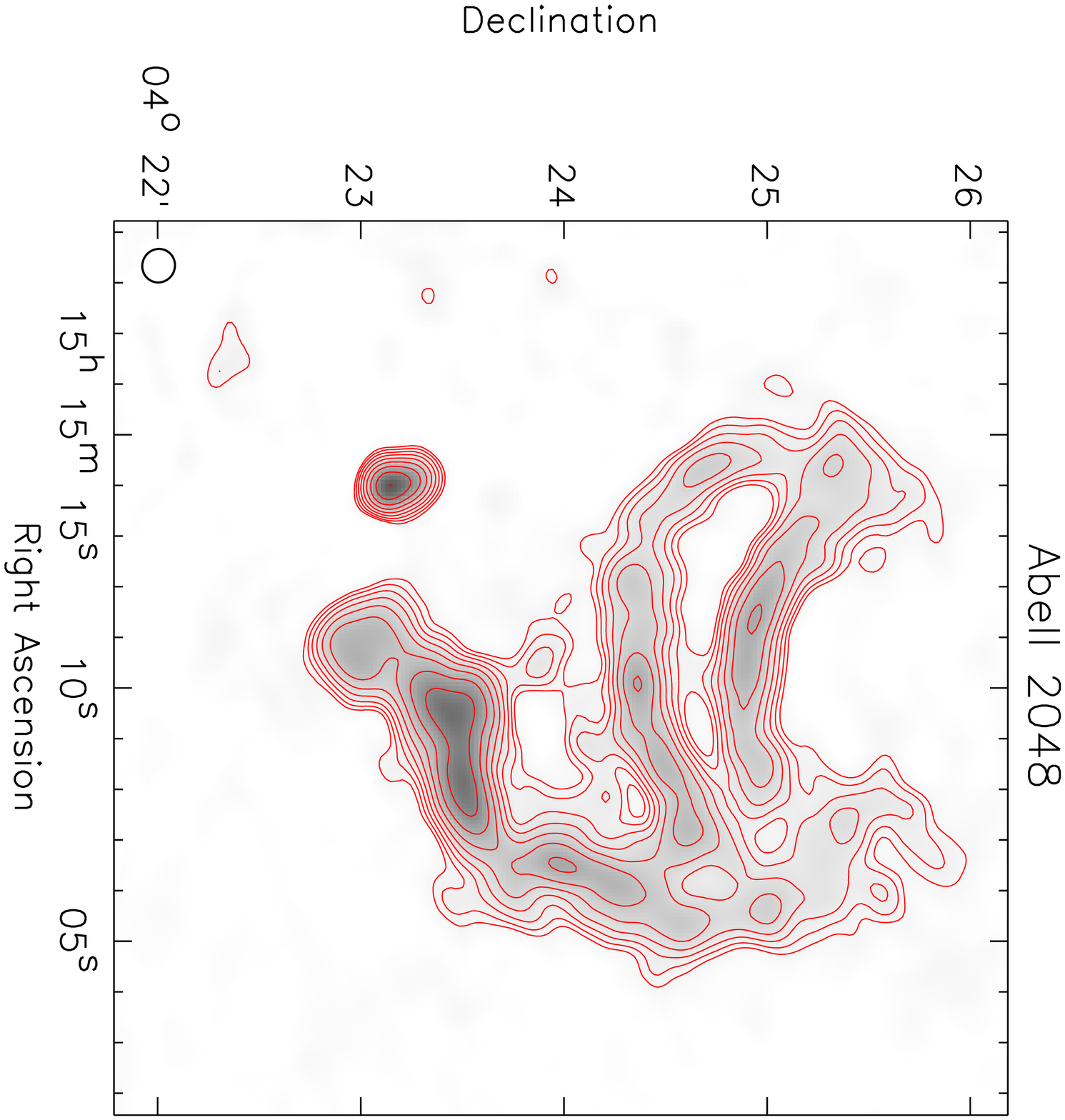}
\includegraphics[angle =90, trim =0cm 0cm 0cm 0cm,width=0.45\textwidth]{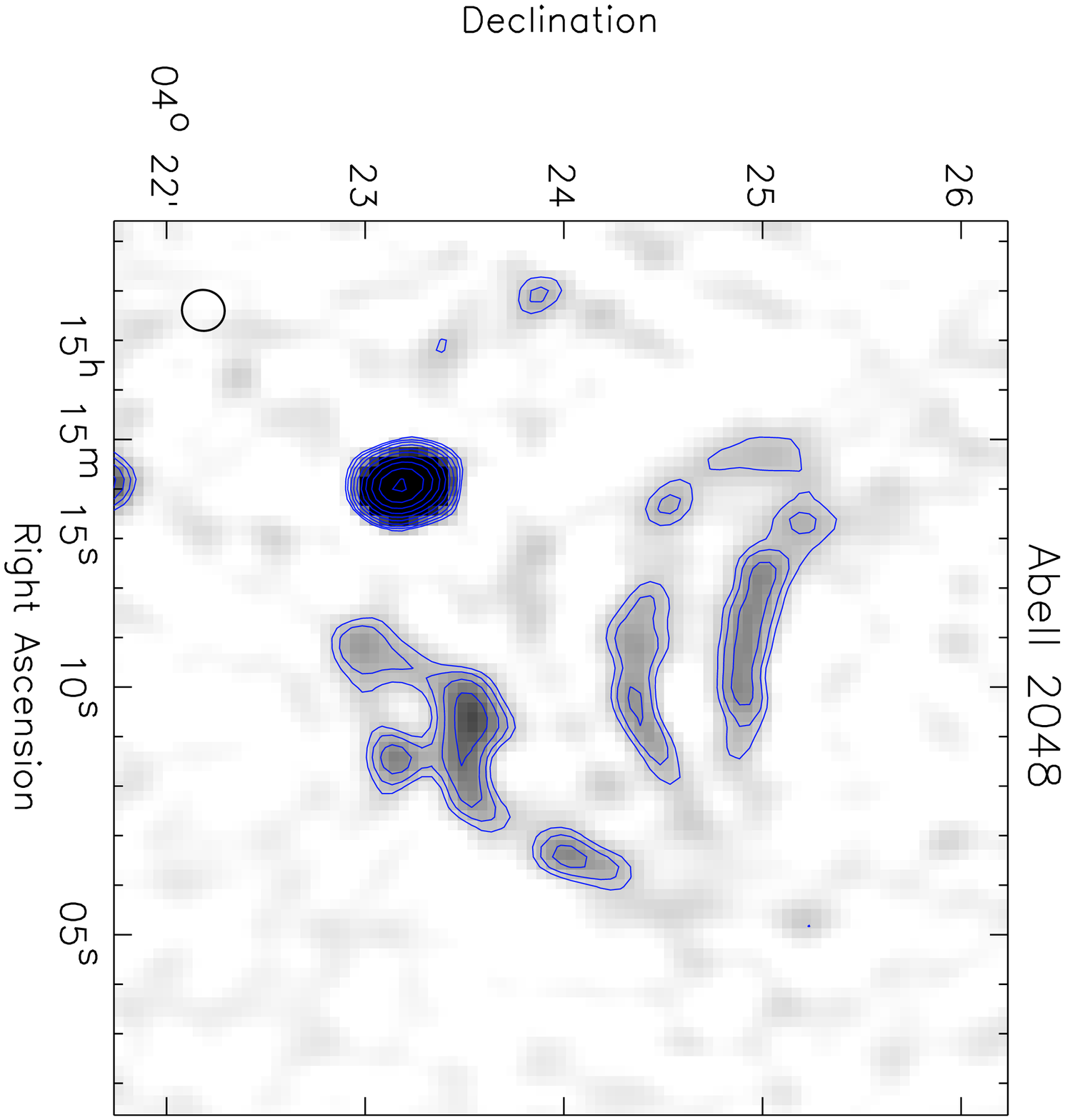}
\includegraphics[angle =90, trim =0cm 0cm 0cm 0cm,width=0.45\textwidth]{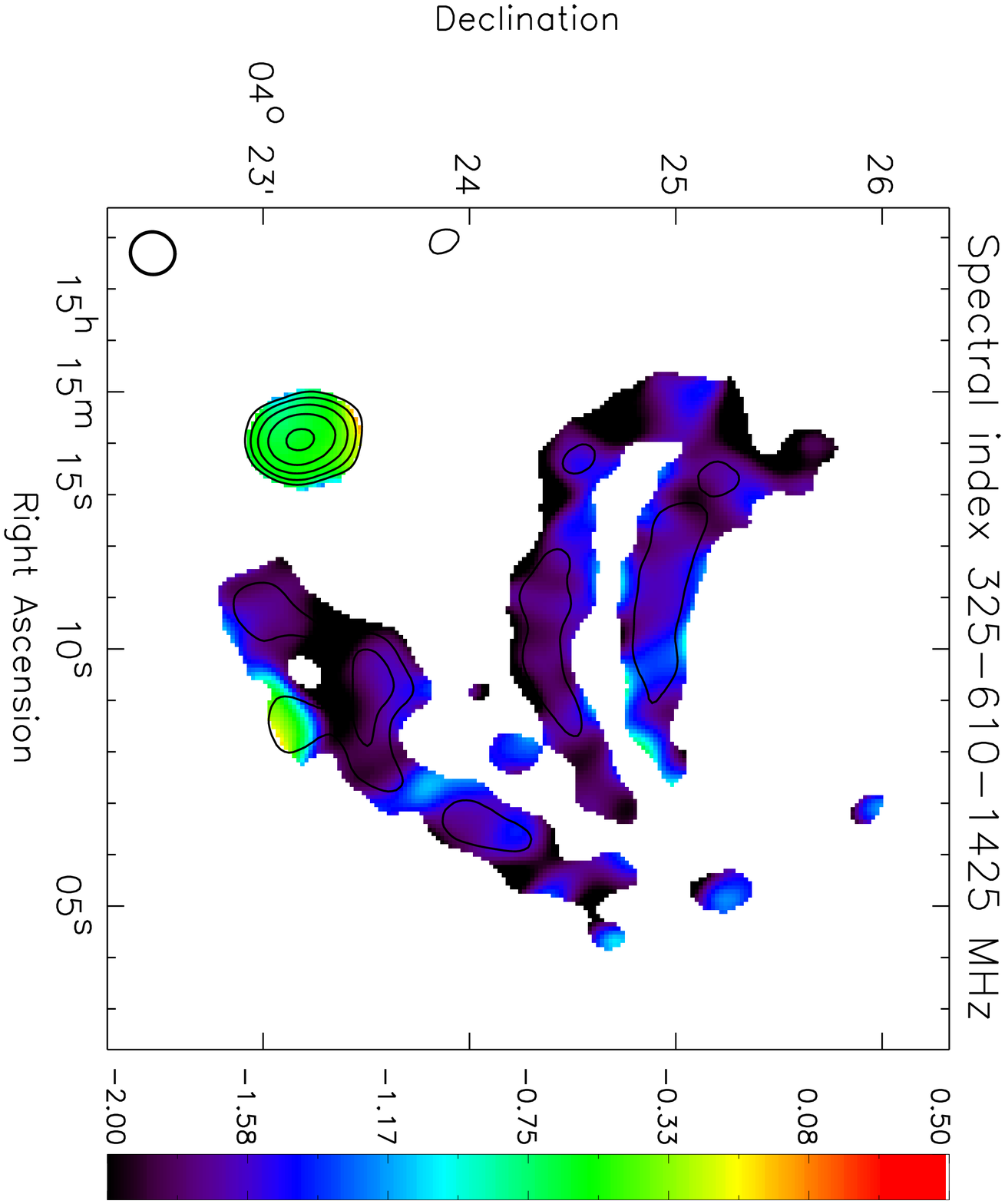}
\includegraphics[angle =90, trim =0cm 0cm 0cm 0cm,width=0.45\textwidth]{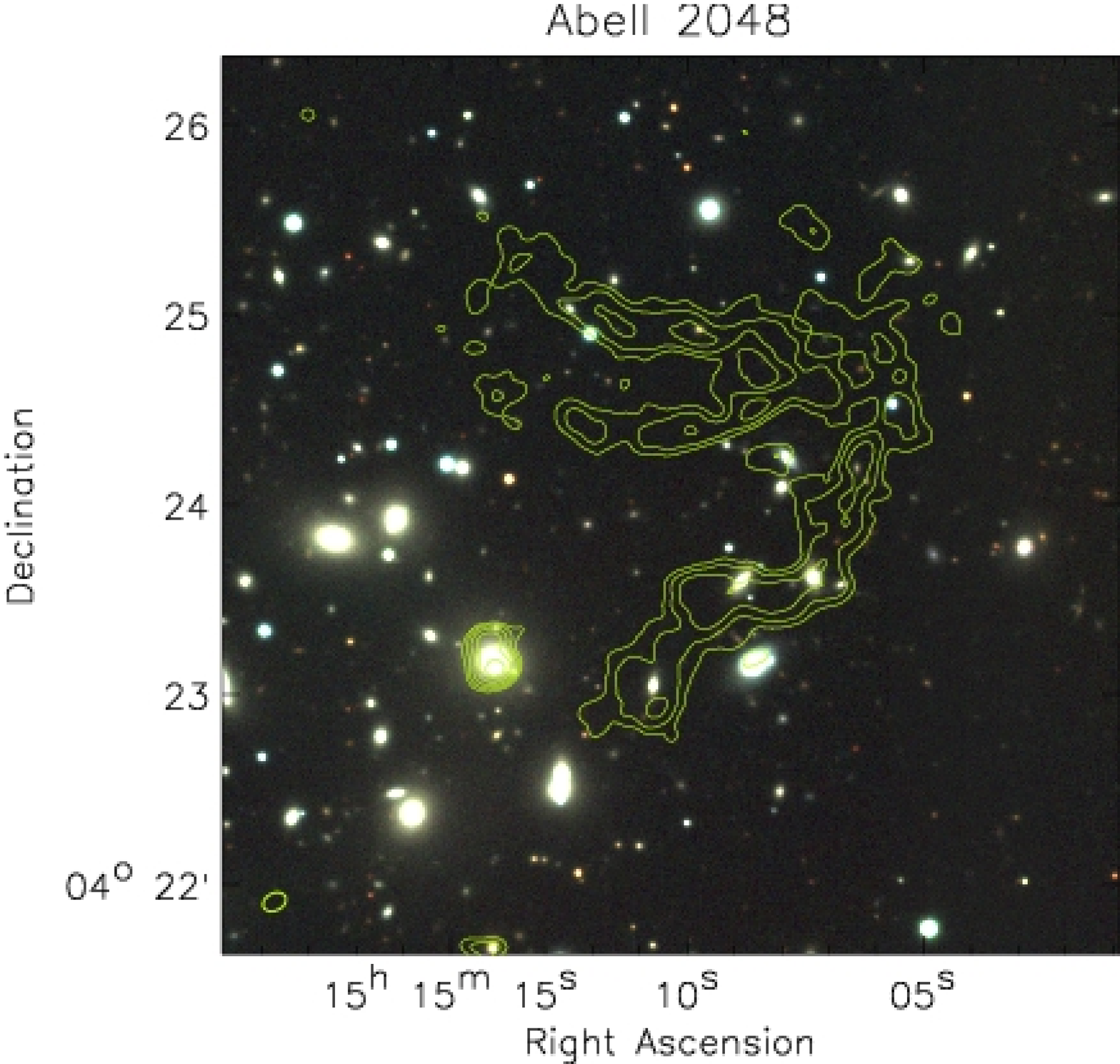}
\end{center}
\caption{Tot left: GMRT 325 MHz map. Contour levels are drawn as in Fig.~\ref{fig:gmrt325_5}. Top right: VLA 1425~MHz map.  Contour levels are drawn as in Fig.~\ref{fig:gmrt325_5}. Bottom left: Power-law spectral index fit between 325, 610, and 1425~MHz. Contours are from the 1.4~GHz VLA image and drawn as in Fig.~\ref{fig:spix5} (left panel). The resolution  is $13.0\arcsec \times 12.4\arcsec$. Bottom right: Optical WHT color image for Abell~2048. GMRT 610~MHz contours are overlaid in yellow. The beam size is $7.6\arcsec \times 5.4\arcsec$. Contour levels are drawn as in Fig.\ref{fig:s6_optical}.}
\label{fig:gmrt325_s25}
\end{figure*}

VLSS~J1515.1+0424 is located in the cluster \object{Abell 2048} \citep[$z=0.0972$; ][]{1999ApJS..125...35S} to the east of the cluster center. The source has a largest extent of 310~kpc (see Fig.~\ref{fig:gmrt325_s25} top left panel), and has a complex morphology. Only the brighter parts of the source are seen in the VLA 1.4~GHz C-array image (Fig.~\ref{fig:gmrt325_s25} top right panel). No polarized flux is detected from the source. We set an an upper limit on the polarization fraction of 8\% for the source, again requiring a SNR of 5 for a detection. An optical V, R, and I color image of the cluster with 610~MHz contours overlaid does not reveal an obvious optical counterpart for the source.

The spectral index map between 325, 610, and 1425~MHz, is shown in Fig.~\ref{fig:gmrt325_s25} (bottom left panel). No systematic spectral index gradients are seen across the source. A region with a flat spectral index ($\alpha > -0.5$) is located under the southern ``arm'' of the source at RA~15$^\mathrm{h}$~15$^\mathrm{m}$~08.6$^\mathrm{s}$, Dec~$+$04\degr~23\arcmin~08\arcsec. This part is associated with the galaxy \object{2MASX~J15150860+0423085} in front of the cluster ($z= 0.047856$ from SDSS DR7) (see Fig.~\ref{fig:gmrt325_s25} bottom right panel). The spectral index of the relic is steep with an average value of about $-1.7$ between 1425 and 325~MHz.

The complex morphology of the radio source suggests that the source can be classified as a radio phoenix. The steep curved radio spectrum is consistent with this interpretation. If the source is indeed a radio phoenix, the radio plasma should have originated in a galaxy that has gone through phases of AGN activity. A candidate is the elliptical galaxy \object{MCG~+01$-$39$-$011} \citep[$z=0.095032$; ][]{1998ApJS..115....1S}. This galaxy is currently active and located close to the eastern end of the southern ``arm''. However, there are several other elliptical galaxies around, although at the moment they are not radio-loud. A ROSAT image \citep[see][]{2009A&A...508...75V} of the cluster shows a substructure to the east of the main cluster, which implies that the cluster is presently undergoing a merger. The velocity dispersion, $\sigma$, of the galaxies in the cluster is 857~km~s$^{-1}$  \citep{2008MNRAS.389.1074S}. The bolometric X-ray luminosity is $1.914 \times 10^{44}$~erg~s$^{-1}$. On the basis of the $L_{\rm{X}}-\sigma$ relation (X-ray luminosity versus velocity dispersion) from \citeauthor{2008MNRAS.389.1074S} 
\begin{equation}
\log{ \left(  \frac{L_{X}}  {10^{44}~\rm{erg~s}^{-1}} \right)   }  = 4.39\log{ \left ( \frac{\sigma}{500~\rm{km~s}^{-1}} \right) }  - 0.530   \mbox{ ,}
\end{equation}
we predict a velocity dispersion of $765\pm40$~km~s$^{-1}$ given the X-ray luminosity. This is lower than the observed value, which is not inconsistent with the cluster having undergone a recent merger event.
A shock wave generated by the proposed merger event might have compressed fossil radio plasma and produced the radio phoenix. Future X-ray observations will be needed to study the dynamical state of the cluster and the relation between the ICM and the radio phoenix.

\section{Optical imaging around five compact steep-spectrum sources}
\label{sec:other}

We present optical images around five slightly more compact steep-spectrum radio 
sources, which nature was found to be unclear in \cite{2009A&A...508...75V}.

\subsection{VLSS J2043.9$-$1118} 
\begin{figure}
\begin{center}
\includegraphics[angle =90, trim =0cm 0cm 0cm 0cm,width=0.5\textwidth]{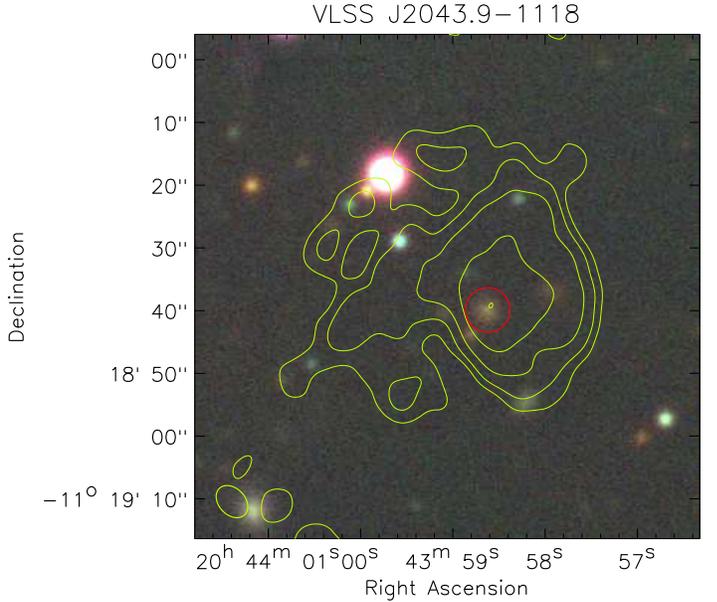}
\end{center}
\caption{Optical WHT color image for VLSS J2043.9$-$1118. GMRT 610~MHz contours are overlaid in yellow. The beam size is $5.8\arcsec \times 4.2\arcsec$. Contour levels are drawn as in Fig.\ref{fig:s6_optical}. A red circle indicates the proposed optical counterpart.}
\label{fig:s10_optical}
\end{figure}

The radio source has a largest angular size of 41\arcsec. An optical counterpart (R band magnitude of 20.1) is visible in our WHT image (see Fig.~\ref{fig:s10_optical}). The radio emission surrounds the galaxy and there is a hint of a faint extension to the east. We estimate a redshift of $0.5 \pm 0.3$ for the optical counterpart, which implies a physical extent of 250~kpc for the radio source. The source could be a mini-halo or core-halo system given its steep spectral index of $-1.74 \pm 0.05$ (between 74 and 1400~MHz). Alternatively, we are detecting  radio plasma from an AGN that has undergone a significant amount of spectral ageing. 

\subsection{VLSS J1117.1+7003} 
This source has a remarkably steep radio spectrum ($\alpha_{74-1400} = -1.87 \pm 0.07$) without any indication of a spectral turnover at low frequencies. The source is resolved into a smooth featureless roughly spherical blob (26\arcsec~by 23\arcsec). We identify a red galaxy, with an R magnitude of $21.2$, as a possible counterpart,  which would put the source at a redshift of $0.8\pm 0.4$. A blue galaxy is located only 5\arcsec~north of the red galaxy.  This might also be the counterpart of the radio source. The integrated R-band magnitude is about the same as the redder galaxy putting it at about the same redshift if it were the optical counterpart. At $z=0.8$, the radio emission would have a physical extent of 130~kpc.  
\begin{figure}
\begin{center}
\includegraphics[angle =90, trim =0cm 0cm 0cm 0cm,width=0.5\textwidth]{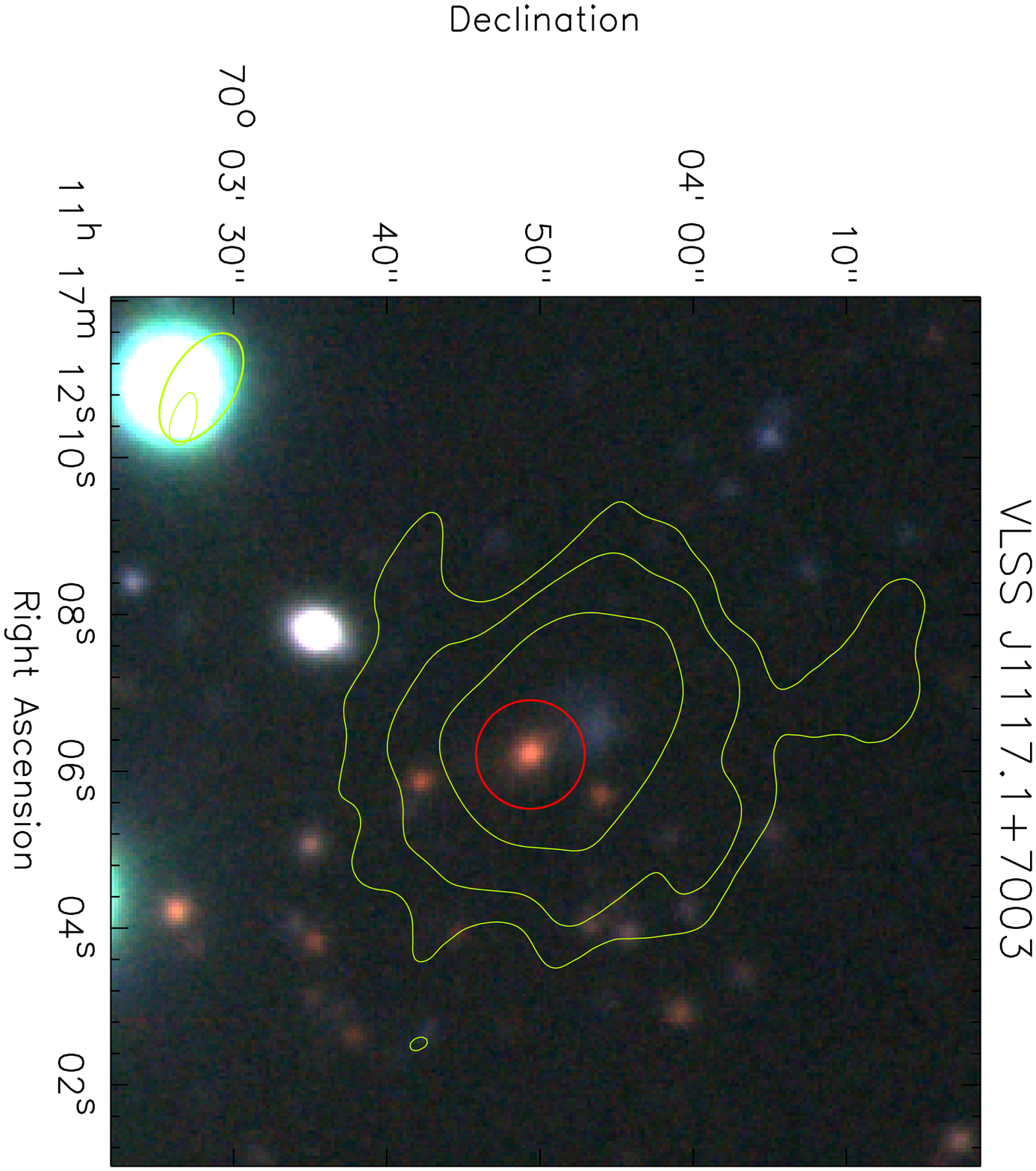}
\end{center}
\caption{Optical WHT color image for VLSS J1117.1+7003. GMRT 610~MHz contours are overlaid in yellow. The beam size is $7.8\arcsec \times 4.3\arcsec$. Contour levels are drawn as in Fig.\ref{fig:s6_optical}. A red circle indicates the proposed optical counterpart.}
\label{fig:s4_optical}
\end{figure}

\subsection{VLSS J2209.5+1546} 
The radio map shows an elongated source. We find a faint (R band magnitude of $22.4$) counterpart halfway along the elongated source. We estimate a redshift of $z=1.1\pm0.7$ (using the Hubble-R relation), giving a physical extent of 500~kpc. 

\begin{figure}
\begin{center}
\includegraphics[angle =90, trim =0cm 0cm 0cm 0cm,width=0.5\textwidth]{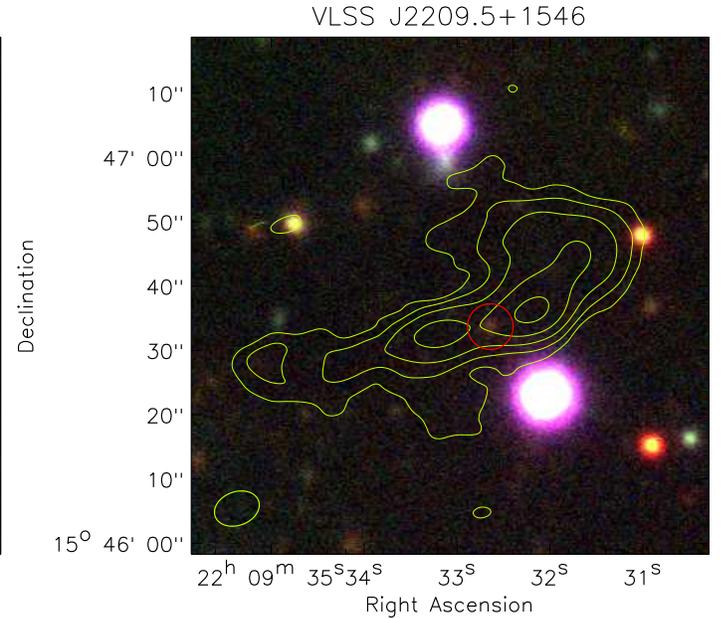}
\end{center}
\caption{Optical WHT color image for VLSS~J2209.5+1546. GMRT 610~MHz contours are overlaid in yellow. The beam size is $6.9\arcsec \times 6.2\arcsec$. Contour levels are drawn as in Fig.\ref{fig:s6_optical}. A red circle indicates the proposed optical counterpart.}
\label{fig:s26_optical}
\end{figure}

\subsection{VLSS J0516.2+0103}   
VLSS~J0516.2+0103 is a slightly elongated source that does not have an optical counterpart in POSS-II images. In our INT image, we identify a possible faint red counterpart with an R magnitude of 22.9. This implies a redshift of $1.2 \pm 0.7$ (including an extinction of 0.367 in the R band), which gives a size of 290~kpc and  makes it a candidate mini-halo or core-halo system. 
\begin{figure}
\begin{center}
\includegraphics[angle =90, trim =0cm 0cm 0cm 0cm,width=0.5\textwidth]{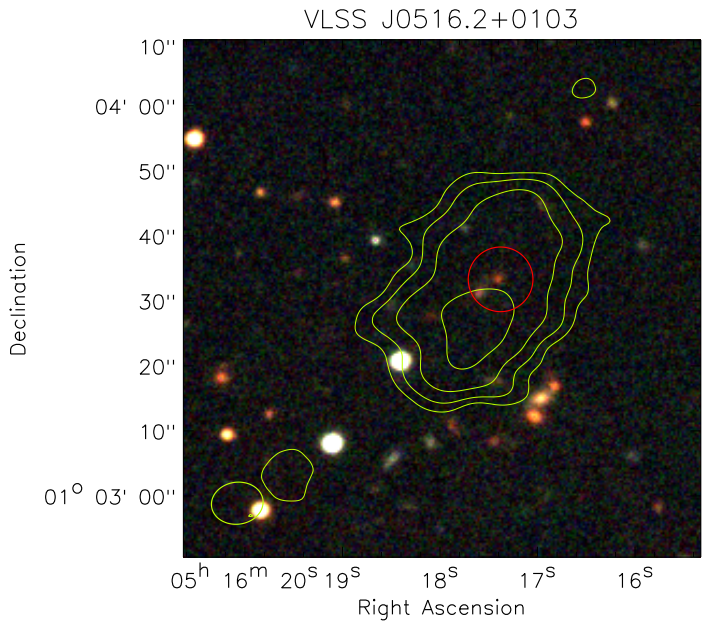}
\end{center}
\caption{Optical INT color image for VLSS~J0516.2+0103. GMRT 610~MHz contours are overlaid in yellow. The beam size is $8.1\arcsec \times 6.5\arcsec$. Contour levels are drawn as in Fig.\ref{fig:s6_optical}. A red circle indicates the proposed optical counterpart.}
\label{fig:s20_optical}
\end{figure}

\subsection{VLSS J2241.3$-$1626}
The morphology of this source is complex. A potential optical counterpart has an R-band magnitude of 20.2 giving a redshift of $0.5 \pm 0.3$ and a physical extent of 290~kpc for the source. The optical counterpart is located roughly halfway along the extended source. The enhancements in the radio emission to the east and west of the proposed counterpart suggest that these are  the lobes of an AGN. The fainter more-extended radio emission might be older radio plasma causing the steep radio spectrum. 
\begin{figure}
\begin{center}
\includegraphics[angle =90, trim =0cm 0cm 0cm 0cm,width=0.5\textwidth]{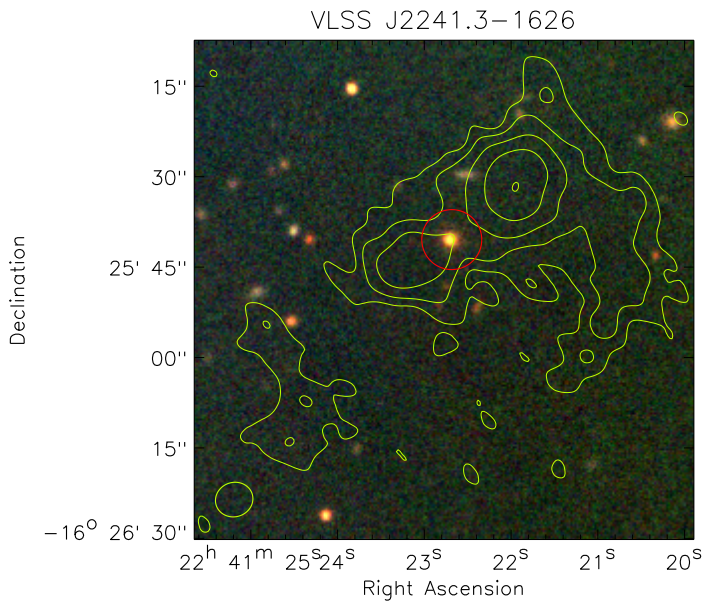}
\end{center}
\caption{Optical INT color image for VLSS~J2241.3$-$1626. GMRT 610~MHz contours are overlaid in yellow. The beam size is $6.1\arcsec \times 5.8\arcsec$. Contour levels are drawn as in Fig.\ref{fig:s6_optical}. A red circle indicates the proposed optical counterpart.}
\label{fig:s28_optical}
\end{figure}

\section{Discussion}
\label{sec:discussion}
Most radio relics and halos known till date are located within massive X-ray luminous clusters. The majority of these sources were discovered in the NVSS and WENSS surveys by visual inspection of the radio maps in and around known galaxy clusters \citep[mostly Abell clusters, ][]{1999NewA....4..141G, 2001ApJ...548..639K}. \cite{2007A&A...463..937V, 2008A&A...484..327V} carried out a search in a complete sample of 50 massive X-ray selected  ($L_{\rm{X,~0.1-2.4~keV}} > 5 \times 10^{44}$~erg~s$^{-1}$) clusters to determine the fraction of radio halos in these systems. The fraction of clusters hosting a giant radio halo was found to be $0.29 \pm 0.09$. The number of small galaxy clusters with low X-ray luminosity known to host a diffuse radio source is very small. An example is the radio halo in \object{Abell 1213} identified by \cite{2009A&A...507.1257G} with $L_{\rm{X,~0.1-2.4~keV}} = 0.1 \times 10^{44}$~erg~s$^{-1}$.

We note that the sources presented in this paper were selected on the basis of their steep spectral index and diffuse nature. There was no requirement for the radio sources to be located in a galaxy cluster. The question arises of whether most radio relics and halos are indeed located in massive galaxy clusters or whether they also occur in poor clusters and galaxy groups. 

None of the sources in our sample, with $\alpha<-1.35$ (between 74 and 1400~MHz),  
are located in massive known galaxy clusters. 
Only VLSS~J1515.1+0424 (in Abell~2048) and VLSS~J1431.8+1331 
(in MaxBCG~J217.95869+13.53470) are found in clusters detected 
in the ROSAT All-Sky Survey. The X-ray luminosities of these clusters 
are moderate, with values between 1--2~$\times 10^{44}$~erg~s$^{-1}$. 
Therefore, our observations indicate that diffuse steep-spectrum 
sources do also occur in less massive clusters and galaxy groups. 
Most of the sources seem to be radio relics related to previous 
episodes of AGN activity: either AGN relics or radio phoenices. 
Some other sources in our sample can be classified as core-halo 
or mini-halo candidates, where the radio emission surrounds a 
central galaxy of a poor cluster or galaxy group. In our case,  
they are not found in massive cool-core clusters.

Amongst the sources in our sample, there are also a number of more distant ($z \sim 1$) filamentary radio sources related to AGN activity. These could be relatively ``nearby'' ultra-steep spectrum (USS) sources \citep[e.g., see ][for a review]{2008A&ARv..15...67M}. As they are relatively nearby, they are clearly extended in for example the 1.4~GHz~FIRST survey images (5\arcsec~resolution) and therefore included in our sample. 

We did not detect any ultra-steep spectrum radio halos \citep{2008Natur.455..944B} in our sample. This could be because the surface brightness of these objects is too low for them to be detected in the 74~MHz VLSS survey \citep{2008Natur.455..944B, 2010A&A...517A..43M}. The 74~MHz VLSS survey is relatively shallow with an average rms noise of 0.1~Jy~beam$^{-1}$.

Since poor galaxy clusters and groups are more numerous, 
it is expected that the sources in our sample are only 
the tip of the iceberg and many more of them should 
turn up in low-frequency surveys, as will be carried out 
for example by LOFAR in the near future. In terms also of the  
timescales related to AGN activity and the ubiquity of 
shocks, these surveys will uncover large populations of AGN 
relics and radio phoenices. One of the difficulties will be 
to classify these sources on the basis of the radio 
morphology, polarization, and spectral index alone. 
The differences between radio phoenices, AGN relics, 
and relics tracing shock fronts with DSA are often subtle. 
The AGN relics and phoenices should have very curved 
radio spectra, while relics caused by electrons accelerated 
at shocks should have straight radio spectra. Nevertheless, deep 
optical/NIR and X-ray surveys will play an important role 
in identifying the nature of these diffuse radio sources.

\section{ Conclusions}
\label{sec:conclusion}
We have presented 325~MHz and 1.4~GHz radio observations of six diffuse steep-spectrum sources. 
The sources were selected from an initial sample of 26 diffuse steep-spectrum ($\alpha < -1.15$) 
sources \citep{2009A&A...508...75V}. Optical WHT and INT images were taken at the positions 
of 10 radio sources from the sample. We briefly summarize the results.
\\
-The radio source VLSS~J1431.8+1331 is located in the cluster  MaxBCG~J217.95869+13.53470 ($z=0.16$)
and associated with the central cD galaxy. A second radio source is located 175~kpc to the east. This source is connected by a faint radio bridge to the central radio source. This source probably traces an old bubble of radio plasma from a previous episode of AGN activity of the central source. The spectral curvature of this source is large, 
indicating the radio plasma is old, which is consistent with the above scenario.
\\
-VLSS J1133.7+2324 is an elongated filamentary steep spectrum source, the nature of the source is unclear. 
It might be a radio relic located in a galaxy cluster at $z \sim 0.6$.
\\
-The relic in Abell~2048 and the source 24P73 are both classified as radio phoenices, 
which consist of compressed 
fossil radio plasma from AGNs. We detect several galaxies close to 24P73, probably 
belonging to the cluster hosting the radio phoenix.
\\
-VLSS J0004.9$-$3457 is a diffuse radio source with emission surrounding the central 
elliptical galaxy of a small cluster or galaxy group. The source could be a radio mini-halo 
(or core-halo system). An arc-like structure is located to the east of the source which 
has a high polarization fraction of 
about 30\% at 1.4~GHz indicative of ordered magnetic fields. This is probably a relic,  where 
either particles are accelerated by the DSA mechanism or radio plasma from the central AGN is compressed.
\\
-The origin of VLSS~J0915.7+2511, a diffuse radio source in MaxBCG~J138.91895+25.19876, 
is somewhat unclear. The source is most likely an AGN relic or radio phoenix.
\\
We also presented optical images around five other diffuse radio source from the sample. 
For these sources, we could not find optical counterparts in POSS-II and 2MASS images. 
We detected candidate counterparts for all of these sources with redshifts in the range $0.5 <  z < 1.2$. 
Some of these sources are radio galaxies, some others may be classified as mini-halos 
as the radio emission surrounds the host galaxy. 

From our observations, we conclude that radio relics are located not 
only in the most massive merging galaxy clusters. 
They can also be found in smaller galaxy clusters and groups. 
Most of these sources probably trace old radio plasma from 
previous episodes of AGN activity. Several other sources resemble mini-halos or core-halos 
that are also found in less massive systems. Future low-frequency 
surveys will probably uncover large numbers of these sources,  
which can then be used to constrain timescales related to AGN 
activity and study the interaction between radio plasma 
and the ICM in clusters and galaxy groups.

\begin{acknowledgements}
We would like to thank the anonymous referee for useful comments. 
We thank the staff of the GMRT who have made 
these observations possible. The GMRT is run 
by the National Centre for Radio Astrophysics of the 
Tata Institute of Fundamental Research. The Westerbork 
Synthesis Radio Telescope is operated by ASTRON 
(Netherlands Institute for Radio Astronomy) with 
support from the Netherlands Foundation for Scientific 
Research (NWO). The National Radio Astronomy 
Observatory is a facility of the National Science 
Foundation operated under cooperative agreement by 
Associated Universities, Inc.  The William Herschel 
Telescope and Isaac Newton Telescope are operated on 
the island of La Palma by the Isaac Newton Group in 
the Spanish Observatorio del Roque de los Muchachos 
of the Instituto de Astrof\'{\i}sica de Canarias. 

This publication makes use of data products from the 
Two Micron All Sky Survey, which is a joint project 
of the University of Massachusetts and the Infrared 
Processing and Analysis Center/California Institute 
of Technology, funded by the National Aeronautics and 
Space Administration and the National Science Foundation. 
This research has made use of the VizieR catalogue 
access tool, CDS, Strasbourg, France. 

The Digitized Sky Surveys were produced at the 
Space Telescope Science Institute under U.S. 
Government grant NAG W-2166. The images of 
these surveys are based on photographic data 
obtained using the Oschin Schmidt Telescope on Palomar 
Mountain and the UK Schmidt Telescope. The plates were 
processed into the present compressed digital form with 
the permission of these institutions. The Second Palomar 
Observatory Sky Survey (POSS-II) was made by the California 
Institute of Technology with funds from the National Science 
Foundation, the National Geographic Society, the Sloan 
Foundation, the Samuel Oschin Foundation, and the Eastman Kodak Corporation.

RJvW would like to thank S.~van~der~Tol for helping with the observations. 
RJvW acknowledges funding from the Royal Netherlands Academy of Arts and Sciences.

\end{acknowledgements}

\bibliographystyle{aa}
\bibliography{15991}

\end{document}